\documentclass{class/jfm}
\usepackage{graphicx}
\usepackage{epstopdf, epsfig}
\usepackage{amsmath}
\usepackage{bm,upgreek}
\usepackage{mathrsfs}
\usepackage[version=4]{mhchem}
\usepackage{siunitx}
\usepackage[abs]{overpic}
\usepackage{xcolor,varwidth}
\usepackage{longtable,tabularx}
\usepackage{upgreek}
\usepackage{setspace}
\usepackage[normalem]{ulem}
\definecolor{orange}{RGB}{255,93,11}
\definecolor{bg}{RGB}{0,0,255}
\definecolor{dg}{RGB}{0,150,0}
\definecolor{lg}{RGB}{46,204,113}
\definecolor{dr}{RGB}{192,57,43}
\usepackage{algorithm}
\usepackage[noend]{algpseudocode}
\newcommand{\B}[1]{\mathbf #1}

\newcommand{\bphi}{\mbox{\boldmath$\phi$}}
\newcommand{\mre}{\mathrm{e}}
\newcommand{\mrd}{\mathrm{d}}
\newcommand{\mri}{\mathrm{i}}
\newcommand{\DefinedAs}[0]{\mathrel{\mathop:}=}
\newcommand{\Ece}{E_{\mbox{\textsc{ce}}}}

\newtheorem{remark}{Remark}

\shorttitle{Reattachment streaks in hypersonic compression ramp flow}
\shortauthor{A.\ Dwivedi, G.\ S.\ Sidharth, J.\ W.\ Nichols, G.\ V.\ Candler, and M.\ R.\ Jovanovi\'c}

\title{\Large Reattachment streaks in hypersonic compression ramp flow: an input-output analysis}

\author{\large Anubhav Dwivedi\aff{1}
  \corresp{\email{dwive016@umn.edu}},
  G.\ S.\ Sidharth\aff{1}, Joseph W.\ Nichols\aff{1}, \\ Graham V.\ Candler\aff{1},
 \and Mihailo R.\ Jovanovi\'c\aff{2}}

\affiliation{\aff{1} Department of Aerospace Engineering and Mechanics, \\ University of Minnesota, Minneapolis, MN 55455, USA
\aff{2}Ming Hsieh Department of Electrical and Computer Engineering, \\ University of Southern California, Los Angeles, CA 90089, USA}

\begin{document}

\maketitle

\begin{abstract}
We employ global input-output analysis to quantify amplification of exogenous disturbances in compressible boundary layer flows. Using the spatial structure of the dominant response to time-periodic inputs, we explain the origin of steady reattachment streaks in a hypersonic flow over a compression ramp. Our analysis of the laminar shock/boundary layer interaction reveals that the streaks arise from a preferential amplification of upstream counter-rotating vortical perturbations with a specific spanwise wavelength. These streaks are associated with heat flux striations at the wall near flow reattachment and they can trigger transition to turbulence. The streak wavelength predicted by our analysis compares favorably with observations from two different hypersonic compression ramp~experiments. Furthermore, our analysis of inviscid transport equations demonstrates that base flow deceleration contributes to the amplification of streamwise velocity and that the baroclinic effects are responsible for the production of streamwise vorticity. Finally, the appearance of the temperature streaks near reattachment is triggered by the growth of streamwise velocity and streamwise vorticity perturbations as well as by the amplification of upstream temperature perturbations by the reattachment shock. 
\end{abstract}

	\vspace*{-5ex}
\section{Introduction}

Compression corners are commonly encountered in intakes, control surfaces, and junctions. High speed flow on a compression corner is a canonical case of shock/boundary layer interaction (SBLI)~\citep{simeonides1995experimental} involving flow separation and reattachment with a shock system. Even though the compression ramp geometry is homogeneous in the spanwise direction, experiments~\citep{re2017experimental} and numerical simulations~\citep{navarro2005numerical} show that the flow over it exhibits three-dimensionality in the form of streamwise streaks near reattachment. The streaks are associated with persistent large local peaks of heat transfer; they can destabilize the boundary layer and cause transition~\citep{simeonides1995experimental,re2017experimental}. 

Recently,~\citet{re2017experimental} and~\citet{chuvakhov2017effect} investigated hypersonic compression ramp flows using temperature sensitive paint (TSP) and infrared (IR) imaging. These techniques were employed to study the formation of streamwise streaks and reattachment heat flux patterns. Previous studies~\citep{inger1977three,simeonides1995experimental,chuvakhov2017effect} attribute the observed structures to G\"ortler-like vortices which develop from upstream perturbations \citep{hall1983linear} and can be amplified by centrifugal effects in the regions of concave streamline curvature near reattachment. However, in most compression ramp studies, the comparison with the theory of G\"ortler instability on curved walls is only qualitative. Furthermore, this theory does not account for the amplification that arises from baroclinic effects in the presence of the wall-normal density gradients~\citep{zapryagaev2013supersonic} and most of the literature neglects the dynamics in the separation bubble. ~\citet{zhuang2017high} used nano-tracer planar laser scattering to visualize a Mach $3$ turbulent boundary layer turning on a $25^{\circ}$ compression ramp. They found that streamwise streaks not only appear after reattachment but also in the separation bubble. It is thus important to understand the role of the recirculation bubble dynamics on the formation and amplification of streamwise streaks.

To include the effect of the separated flow, \citet{sidharth2018onset} carried out a global stability analysis and discovered a 3D global instability in the separation bubble, which results in temperature streaks post-reattachment. The spanwise wavelength of the global instability scales with the recirculation length \citep{gs2017global}. This is in contrast to the spanwise wavelength observed for reattachment streaks~\citep{chuvakhov2017effect,re2017experimental,navarro2005numerical}, which scale with the separated boundary layer thickness, indicating that the global instability is not responsible for their formation. To characterize the role of external perturbations in the formation of these streaks, we consider compression ramp flows that do not exhibit 3D global instability. External disturbances are amplified as they pass through the flow field and we utilize global input-output (I/O) analysis to quantify this amplification. 

\begin{figure}
\centering
\includegraphics[width=0.65\linewidth]{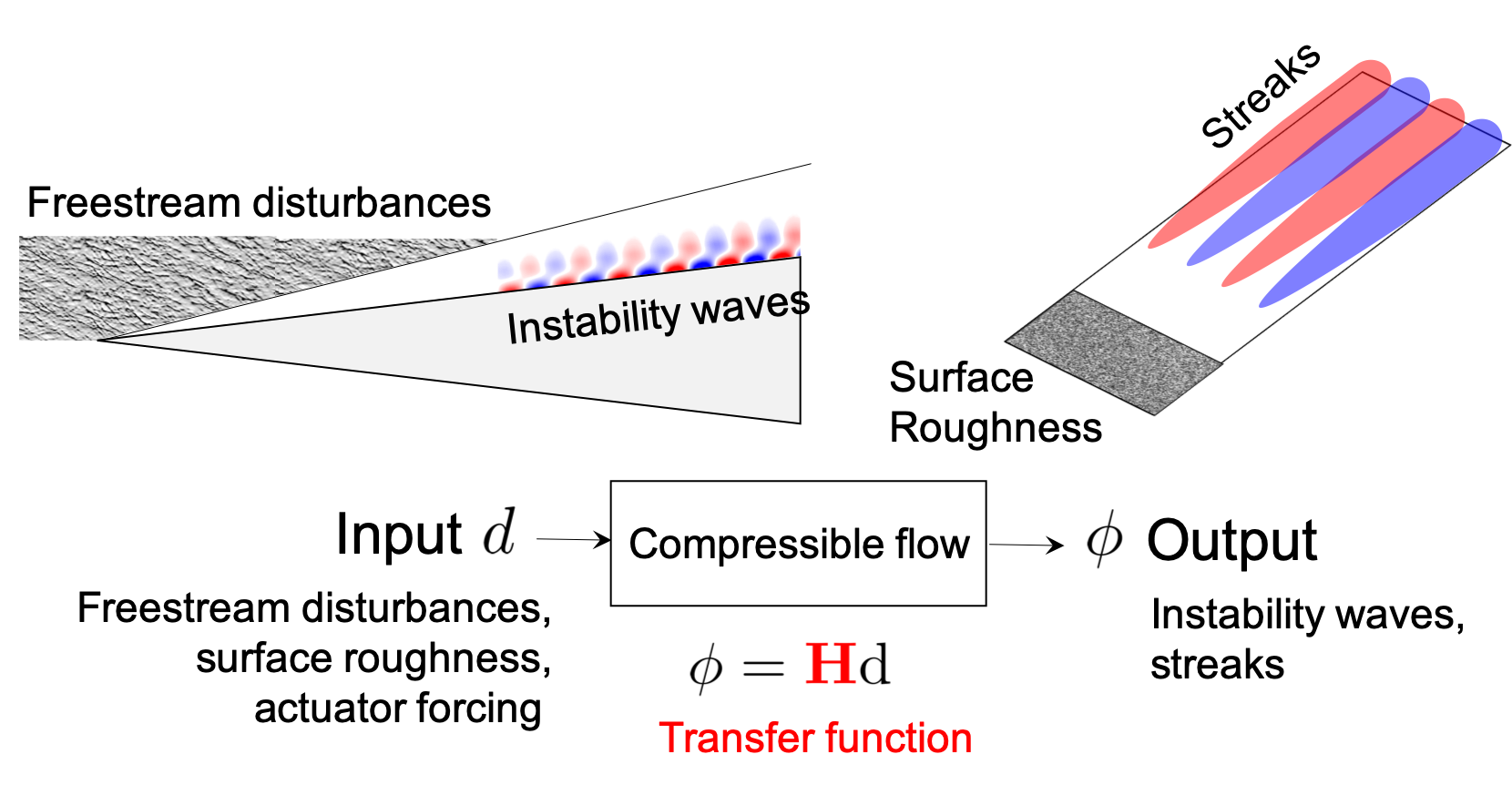}%fig1_enlarged.png}
\caption{Schematic of the input-output approach to compressible flow instabilities.}
\label{fig:fig1}
\end{figure}

The I/O analysis evaluates the response (outputs) of a dynamical system to external perturbation sources (inputs). For time-periodic inputs, the transfer function maps the input forcing to output responses; see~figure~\ref{fig:fig1} for an illustration. For small perturbations, the transfer function can be obtained by linearizing the compressible Navier-Stokes (NS) equations around a laminar base flow. The I/O approach has been employed to quantify amplification and study  transition mechanisms in channels~\citep{mj-phd04,jovanovic2005componentwise}, boundary layers~\citep{brandt2011effect,sipp2013characterization,fosas2017optimal,ranzarhacjovPRF18,JWN2018FS}, and jets~\citep{jeun2016input,schmidt2018spectral}. 

In this paper, we utilize the I/O analysis to demonstrate that the hypersonic shock/boundary layer interaction over a compression ramp strongly amplifies low-frequency upstream disturbances with a specific spanwise length scale. The dominant I/O pair resulting from our analysis is used to explain the emergence of reattachment streaks and to compare our results with experiments. We utilize direct numerical simulations (DNS) to verify the presence of reattachment streaks in the flow subject to dominant steady and unsteady inputs. To uncover physical mechanisms responsible for streak amplification, we also conduct the analysis of inviscid transport equations associated with velocity, vorticity, and temperature perturbations. We show that base flow deceleration contributes to the amplification of streamwise velocity and that the baroclinic effects are responsible for the amplification of streamwise vorticity. Furthermore, the appearance of the temperature streaks near reattachment is triggered by the growth of streamwise velocity and vorticity as well as by the amplification of upstream temperature perturbations by the reattachment shock. In contrast to previous studies~\citep{chuvakhov2017effect,re2017experimental,navarro2005numerical}, our analysis demonstrates the importance of baroclinic terms in cold wall hypersonic boundary layers and shows that the centrifugal effects play only a minor role in the emergence of steady reattachment streaks. We also show that the spanwise scale selection results from the interplay between the presence of flow perturbations in the separation bubble and in the \mbox{reattaching shear layer.}

Our presentation is organized as follows. In \S~\ref{sec.background}, we present the linearized model and provide a brief summary of the I/O formulation. We compute the amplification in attached supersonic flat-plate boundary layers and verify our method against state-of-the-art approaches. In \S~\ref{sec.main}, we evaluate the frequency response of 2D laminar hypersonic base flow on a compression ramp to 3D upstream disturbances and illustrate that the dominant output field appears in the form of steady streamwise streaks near reattachment. We verify the robustness of the dominant response predicted by our analysis using DNS and visualize its spatial structure to illustrate the role of various flow regions in perturbation amplification. In \S~\ref{sec.physics}, we examine inviscid transport equations, investigate production of flow perturbations by the base flow gradients, and uncover physical mechanisms driving the growth of reattachment streaks. We conclude our presentation in \S~\ref{sec.remarks}. 

	\vspace*{-3ex}
\section{Input-output formulation for compressible flows}
	\label{sec.background}

The compressible NS equations for perfect gas in conservative form are given by
\begin{equation}
\label{eq:NSE}
\frac{\partial\mathbf{U}}{\partial t} \; + \; \frac{\partial \mathbf{F}_{j}}{\partial x_{j}} \; = \; 0, 
\end{equation}
where $\mathbf{F}_{j}(\B U)$ is the flux vector and $\mathbf{U}=\big (\rho, \rho \mathbf{u}, E\big )$ is the vector of conserved variables representing mass, momentum, and total energy per unit volume of the gas~\citep{candler2015development}. We decompose the state vector $\B U(\B x,t)$ into a steady base component $\overline{\B U}(\B x)$ and a time-varying perturbation component $\B U'(\B x, t)$, $\B U(\B x,t) = \overline{\B U}(\B x) + \B U'(\B x, t)$. The evolution of small perturbations is then governed by the linearized flow equations, 
	\begin{align}
\frac{\partial}{\partial t} \, \mathbf{U}'(\mathbf{x},t) & \; = \; \mathcal{A}(\overline{\B U}) \mathbf{U}'(\mathbf{x},t), 
	\label{eq:lin}
	\end{align}
where $\mathcal{A}(\overline{\mathbf{U}})$ represents the compressible NS operator resulting from linearization of~\eqref{eq:NSE} around the base flow $\overline{\B U}$. A second order central finite volume discretization \citep{sidharth2018onset} is used to obtain the finite dimensional approximation of Eq.~\eqref{eq:lin},
	\begin{align}
	\frac{\mrd}{\mrd t} \, \mathbf{q} & \; = \; \B A \B q,
	\label{eq:SDlin}
	\end{align}
which describes the dynamics of the spatially discretized perturbation vector $\B q$.

In this paper, we are interested in quantifying the amplification of exogenous disturbances in boundary layer flows~\citep{jovanovic2005componentwise,schmid2007nonmodal}. To accomplish this objective, we augment the evolution model~\eqref{eq:SDlin} with external excitation sources
\begin{align}
\begin{split}
\label{eq:SScomp}
\frac{\mrd}{\mrd t} \, \mathbf{q} & \; = \; \B A \B q \; + \; \B B \B d, \\
\bphi & \; = \; \B C \B q,
\end{split}
\end{align}
where $\mathbf{d}$ is a spatially distributed and temporally varying disturbance source (input) and $\bphi = ( \rho', {\B u}', T' )$ is the quantity of interest (output), where $T'$ denotes temperature perturbations. In Eq.~\eqref{eq:SScomp} the matrix $\B B$ specifies how the input enters into the state equation, while the matrix $\B C$ extracts the output from the state $\B q$. An I/O relation is obtained by applying the Laplace transform to~\eqref{eq:SScomp}, 
\begin{align}
    \label{eq:SSlap}
    \bphi(s) \; = \; \B C(s\B I \, - \, \B A)^{-1}( \B q(0) \, + \, \B B \B d(s)),
\end{align}
where $\B q(0)$ denotes the initial condition and $s$ is the complex number. Equation~\eqref{eq:SSlap} can be used to characterize both the unforced (to initial condition ${\B q}(0)$) and forced (to external disturbances ${\B d}$) responses of the flow perturbations. 

In boundary layer flows, the linearized flow system is globally stable. Thus, for a time-periodic input with frequency $\omega$, $\B d (t) = \hat{\B d} (\omega) \mathrm{e}^{\mri \omega t}$, the steady-state output of a stable system~\eqref{eq:SScomp} is given by $\bphi (t) = \hat{\bphi} (\omega) \mre^{\mri \omega t}$, where $\hat{\bphi} (\omega) = \B H (\mri \omega) \hat{\B d} (\omega)$ and $\B H (\mri \omega)$ is the frequency response 
\begin{align}
    \label{eq:output}
    \B H (\mri \omega) & \; = \; \B C(\mri \omega \B I \, - \, \B A)^{-1} \B B.
\end{align}
At any $\omega$, the singular value decomposition of $ \B H (\mri \omega)$ can be used to quantify amplification of time-periodic inputs~\citep{mj-phd04,schmid2007nonmodal,mckeon2010critical}, 
\begin{align}
    \label{eq:svd}
          \B H (\mri \omega) \B D (\mri \omega) 
          \; = \; 
          \bm{\Upphi} (\mri \omega) \bm{ \Upsigma} (\mri \omega)   
          \quad \Leftrightarrow \quad 
          {\B H } (\mri \omega) 
          \; = \; 
          \bm{\Upphi} (\mri \omega) \bm{\Upsigma} (\mri \omega) {\B D}^{*} (\mri \omega).
\end{align}
Here, $(\cdot)^{*}$ denotes the complex-conjugate transpose, $\bm{\Upphi}$ and $\B D$ are unitary matrices, and $\bm{\Upsigma}$ is the rectangular diagonal matrix of the singular values $\sigma_i (\omega)$. The columns $\mathbf{d}_{i}$ of the matrix $\B D$ represent the input forcing directions that are mapped through the frequency response $\B H$ to the corresponding columns $\bphi_{i}$ of the matrix $\bm{\Upphi}$; for $\hat{\B d} = \mathbf{d}_i$, the output $\hat{\bphi}$ is in the direction $\bphi_i$ and the amplification is determined by the corresponding singular value $\sigma_{i}$. For a given temporal frequency $\omega$, we use a matrix-free approach~\citep{dwivedi2018input} to compute the largest singular value $\sigma_1 (\omega)$ of $
\B H (\mri \omega)$. Note that, at any $\omega$, 
	\begin{equation}
	\text{G} (\omega) 
	\; \DefinedAs \; 
	\sigma_{1} (\omega) 
	\; = \; 
	\dfrac{\| \B H (\mri \omega) \mathbf{d}_{1} (\omega)\|_{E}}{\| \mathbf{d}_{1} (\omega)\|_{E}}
	\; = \; 
	\dfrac{\| \bphi_{1} (\omega) \|_{E}}{\| \mathbf{d}_{1} (\omega)\|_{E}},
	\label{eq.G}
	\end{equation}
\noindent denotes the largest induced gain with respect to Chu's compressible energy norm~\citep{hanifi1996transient}, where ($\mathbf{d}_1 (\omega),\bphi_1 (\omega)$) identify the spatial structure of the dominant I/O pair.

	\vspace*{-1ex}
 \subsection{Validation: supersonic flat plate boundary layer}
 \label{sec:BLAnalysis}
 
Before analyzing the amplification of disturbances in a hypersonic flow involving shock/boundary layer interaction, we apply I/O analysis to compute amplification in a supersonic flow over a flat plate. Our computations are verified against conventional approaches to demonstrate the agreement for canonical problems. Two amplification mechanisms are considered: two-dimensional unsteady acoustic amplification \citep{ma2003receptivity} and three-dimensional steady lift-up amplification \citep{zuccher2005optimal}. 

%	\vspace*{-1ex}
\subsubsection{Two-dimensional unsteady perturbations: acoustic amplification}

\begin{figure}
\centering
\includegraphics[width=0.75\linewidth]{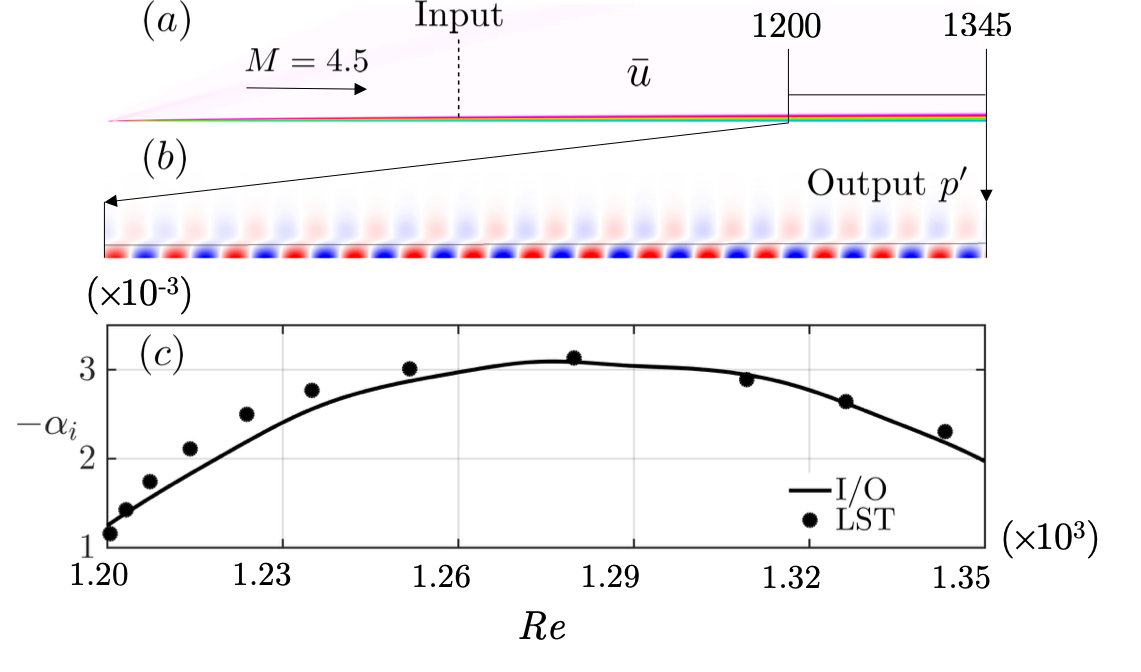}%{acoustic_nov9.png}
\caption{(a) Results of I/O analysis applied to Mach $4.5$ adiabatic boundary layer with perturbation input at $Re=850$; (b) pressure perturbations corresponding to the principal output direction $\bphi_1$; and (c) comparison of spatial growth rate with LST~\citep{ma2003receptivity}.}
\label{fig:fig2}
\end{figure}

% note: non-dimensional $F = 1.6 \times 10^{-4}$} equals to $f=133\, \mbox{KHz}$
Local spatial instabilities corresponding to acoustic perturbations dominate the transition in high speed flat plate boundary layers~\citep{fedorov2011transition}. Using local spatial linear stability theory (LST) and direct numerical simulations (DNS),~\citet{ma2003receptivity} showed that perturbation with non-dimensional frequencies $0.6 \times 10^{-4} <F< 2.2 \times 10^{-4}$ result in spatial growth due to the local instability over a part of the domain. We consider I/O analysis at $F = 1.6 \times 10^{-4}$ and compare the region of growth with that predicted by LST~\citep{ma2003receptivity}; see figure~\ref{fig:fig2}(a) for geometry. The base flow is computed using the finite volume compressible flow solver US3D~\citep{candler2015development} with $125$ cells in the wall-normal and $1600$ cells in the streamwise direction. This resolution yields grid-insensitive I/O results.

% $Re =850$ corresponds to $x_{0}=0.1 \,\mbox{m}$ downstream of the leading edge. 
% $Re = 1300$ corresponds to $x=0.18 \,\mbox{m}$

As shown in figure~\ref{fig:fig2}(a), we use the matrix $\B B$ in Eq.~\eqref{eq:SScomp} to localize the disturbance input at a streamwise location corresponding to the local Reynolds number $Re\DefinedAs\sqrt{Re_{x}} =850$, where the Reynolds number $Re_{x}$ is based on the distance $x$ downstream of the leading edge. This choice allows us to avoid large streamwise gradients in the base flow in the vicinity of the leading edge. The slow streamwise variation of the base flow implies that LST is approximately valid downstream of this location. Furthermore, this location is sufficiently upstream of the neutral point of the acoustic instability, which takes place at $Re = 1140$. This ensures that any non-modal growth arising from the Orr-mechanism~\citep{dwivedi2018input} decays before the spatial growth rate of the local acoustic instability becomes positive. Sponge regions are used at the top and right boundaries to model non-reflecting radiation boundary conditions. We have verified independence of our results on the strength and the location of the sponge zones.

The output of interest is chosen to be the perturbation field in the entire domain, i.e., $\bphi = \B q$. Figure~\ref{fig:fig2}(b) shows the spatial structure of pressure perturbation in the principal output mode $ \bphi_1$. We compute the local spatial growth rate from pressure at the wall $\hat p_\mathrm{wall}$, $\alpha_{i} =-( \partial \hat{ p}_\mathrm{wall}/\partial x) /\hat{p}_\mathrm{wall}$. Figure~\ref{fig:fig2}(c) shows that our I/O analysis correctly identifies the region of spatial instability and predicts growth rates that are close to those resulting from LST~\citep{ma2003receptivity}.  The difference can be attributed to the fact that LST does not account for the spatially-growing nature of the base flow.

	\vspace*{-1ex}
\subsubsection{Three-dimensional steady perturbations: lift-up mechanism}

The spatially-developing boundary layer also supports significant growth of perturbations that are not related to a dominant eigenmode of the linearized dynamical generator. For example, the steady 3D streak-like perturbations that result from the lift-up mechanism~\citep{ellingsen1975stability} play an important role in transition induced by distributed surface roughness~\citep{reshotko2001transient}. \cite{zuccher2005optimal}~used the linearized boundary layer (BL) equations to compute spatial transient growth and analyze this mechanism. For verification purposes, we compare the spanwise wavelength of the maximally amplified streaks resulting from the linearized BL equations and the I/O analysis. We specifically consider the conditions in~\citet{zuccher2005optimal} corresponding to a boundary layer on a 2D adiabatic flat plate in a supersonic free-stream.

\begin{figure}
\centering
\includegraphics[width=0.99\linewidth]{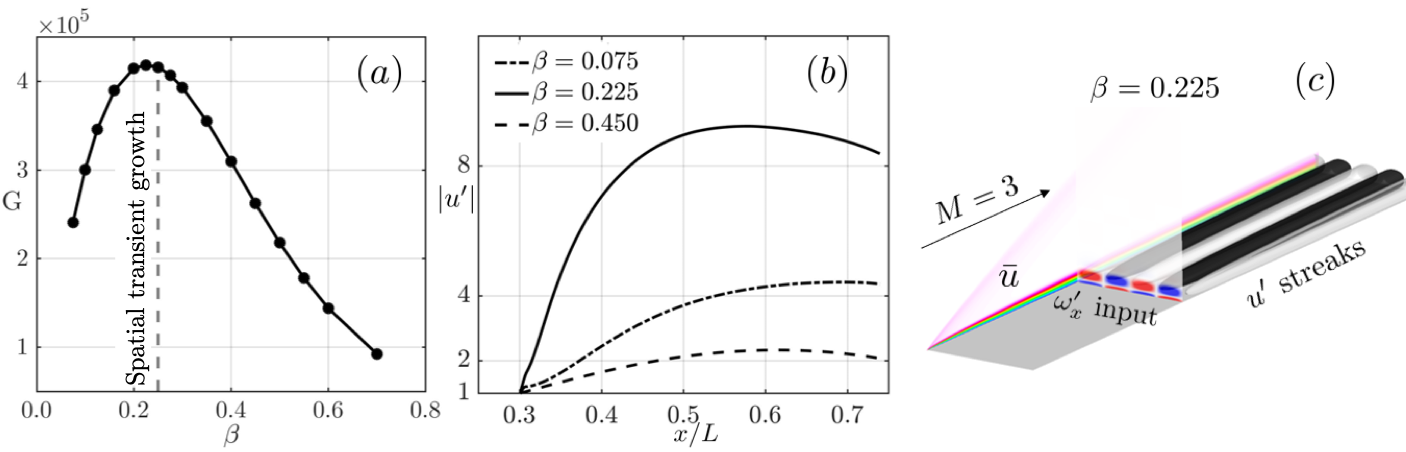}
\caption{(a) Optimal I/O amplification for steady perturbations; (b) streamwise velocity perturbation for different spanwise wavenumbers $\beta$ along the base-flow streamlines; and (c) contours of streamwise vorticity input ($\mathbf{d}_{1}$) and isosurfaces of streamwise velocity output ($\bphi_{1}$) along the BL (shown using base flow streamwise velocity). The dashed curve in (a) indicates the spanwise wavenumber from spatial transient growth calculations~\citep{zuccher2005optimal}.}.
\label{fig:fig3}
\end{figure}

A grid with $250$ cells in the wall-normal and $600$ cells in the streamwise direction is used to compute the base flow and conduct I/O analysis. The input is localized to a plane at streamwise location $x/L=0.3$, where $L$ denotes the plate length. As in the previous subsection, the output is the perturbation field in the entire domain. Owing to homogeneity in the spanwise direction, 3D perturbations take the form $\B q (x,y,z,t) = \tilde{\B q}(x,y) \mre^{\mri(\beta z - \omega t)} $, where $\beta=2\pi/\lambda_z$ is the spanwise wavenumber. Here, the spanwise and wall-normal coordinates are non-dimensionalized using the viscous length scale, $L/Re_{L}$, where $Re_{L}$ is the Reynolds number based on the plate length $L$. To capture the steady lift-up mechanism, we conduct the I/O analysis for $\omega=0$. In figure~\ref{fig:fig3}(a), we illustrate the $\beta$-dependence of the gain $\text{G}$ resulting from the I/O analysis and identify the value of $\beta$ at which the largest spatial transient growth takes place. This value is slightly smaller than the one reported in~\cite{zuccher2005optimal}. We attribute the observed mismatch to different base flow profiles; while we use a numerically computed 2D base flow,~\cite{zuccher2005optimal} used an analytical self-similar base flow profile. The input $\mathbf{d}_1$ (shown in figure~\ref{fig:fig3}(c)) consists of streamwise vortical perturbations and the output $\bphi_1$ consists of a rapid development of streamwise velocity streaks. The algebraic nature of the growth (as opposed to exponential) is illustrated in figure~\ref{fig:fig3}(b) for different $\beta$. As expected, a large initial transient growth is followed by eventual downstream decay. 

The above results show that I/O analysis correctly captures the physical mechanisms responsible for amplification in canonical supersonic flows. As we demonstrate in the next section, this analysis also provides useful insight about the early stages of transition in complex hypersonic compression ramp flow with shock/boundary layer interaction. 

	\vspace*{-2ex}
\section{Input-output analysis of hypersonic compression ramp flow}
	\label{sec.main}

Streamwise streaks in wall temperature are often observed in compression ramp experiments. Although their appearance is typically attributed to amplification that arises near reattachment from centrifugal~\citep{navarro2005numerical,chuvakhov2017effect} or baroclinic effects~\citep{zapryagaev2013supersonic}, quantifying amplification in the presence of a recirculation bubble is an open challenge. Herein, we employ the I/O framework to study the amplification of infinitesimal spanwise periodic upstream disturbances in hypersonic compression ramp flow and explain origin of the heat streaks at reattachment.

Recently,~\citet{re2017experimental} and~\citet{chuvakhov2017effect} reported multiple hypersonic compression ramp experiments in two different facilities with matched free-stream Mach and Reynolds numbers. Temperature sensitive paint (TSP) and infrared thermography measurements of reattachment heat-flux wall patterns revealed quantitatively similar streaks. The effects of free-stream Reynolds number and leading edge radius on the spanwise wavelength $\lambda_z$ of the streaks were also reported. Our objective is to identify the streak wavelength $\lambda_z$ that is selected by the linearized compressible NS equations in the shock/boundary layer interaction.
 
\begin{figure}
\centering
\includegraphics[width=0.99\linewidth]{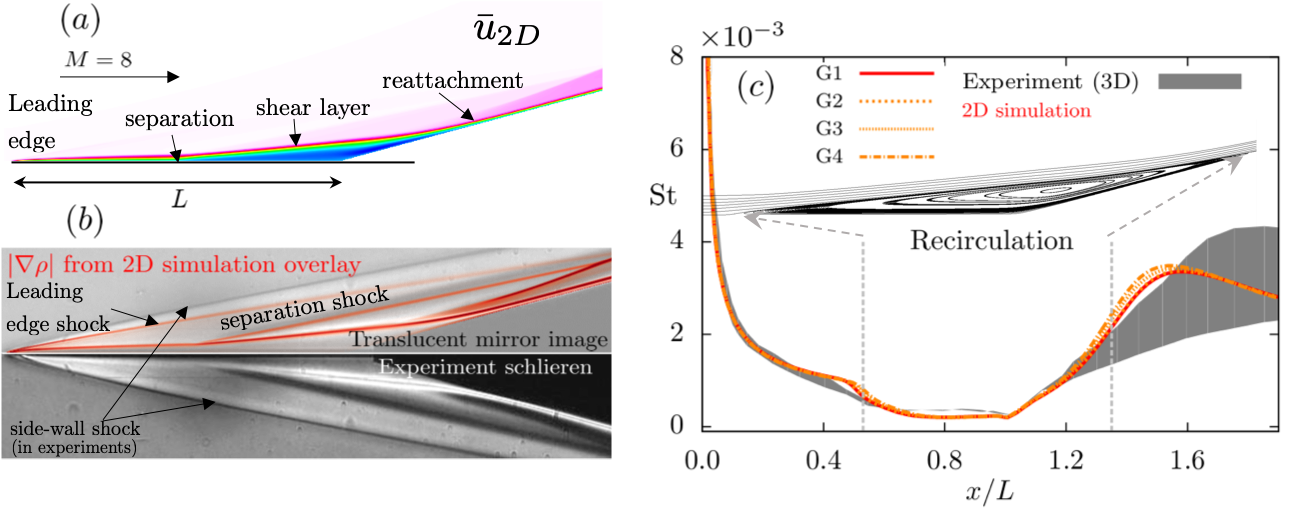}%{base_apr8.png}
\caption{(a) Flow geometry and 2D steady streamwise velocity at $Re_{L}=3.7\times 10^5$; (b) comparison to experimental schlieren; and (c) variation of Stanton number ($St$) with $x/L$. 
Curves G1-G4 denote computational grids at varying resolution ($n_\xi \times n_\eta$), with G1~($577\times349$), G2~($495\times300$), G3~($412\times 249$), and G4~($330\times200$) where $n_\xi$ and $n_\eta$ denote the number of streamwise and wall normal grid points, respectively. The shaded grey region denotes envelope of spanwise variation of $St$ measured in experiments.}
\label{fig:base}
\end{figure}

We consider the experiments performed in the UT-1M Ludwig tube~\citep{chuvakhov2017effect} at Mach 8 with a test time  $T_{\mathrm{test}} = 40\,\mbox{ms}$. As illustrated in figure~\ref{fig:base}(a), the geometry consists of an $L=50 \,\mbox{mm}$ isothermal flat plate with a sharp leading edge and wall temperature $T_w=293 \,\mbox{K}$, followed by an inclined ramp at $15^{\circ}$. The streamwise domain extends from $x/L=0$ to $x/L=1.65$. Table~\ref{tab:fs} summarizes the two free-stream conditions that are considered in our study. We note that the aforementioned test time is large compared to the convective time-scale $L/U_\infty$, $T_\mathrm{test} = 1000L/U_\infty$. Figure~\ref{fig:base}(b) provides comparison of the experimental schlieren image with the 2D base flow density gradient magnitude field that we computed using US3D. Our 2D simulations correctly capture the presence of both the separation and reattachment shocks. The mismatch {near the leading edge} is attributed to the presence of {strong oblique} shocks that originate from the side-walls which are required to maintain 2D flow in experiments but are completely absent in numerically computed 2D base flow. {As seen from the computed flow field, the corresponding shock from the sharp leading edge is significantly weaker and is not captured clearly in the experimental schlieren.}

\begin{table}
\centering
\begin{tabular}{ c c c c c }
  $Re_L$ &$p_\infty$ & $T_\infty$ & $U_\infty$ & $\rho_\infty$\\ \hline
  $3.7\times10^{5}$ & $355 \mbox{ Pa}$& $55 \mbox{ K}$ &$1190 \mbox{ m/s}$ & $0.022 \mbox{ Kg/m}^{3}$\\
  $2.0\times10^{5}$ &$ 164 \mbox{ Pa}$& $55 \mbox{ K} $&$1188 \mbox{ m/s}$ & $0.010 \mbox{ Kg/m}^{3}$\\
\end{tabular}
\caption{Free-stream conditions for experiments reported in \citet{chuvakhov2017effect} and \citet{re2017experimental}, respectively. Reynolds number $Re_{L}$ is based on the plate length $L$.}
\label{tab:fs}
\end{table}

\indent {The Stanton number $St$ is a non-dimensional parameter that determines the wall heat-transfer coefficient~\citep{schlichting2016boundary,chuvakhov2017effect},
\begin{align}
St \; = \; \dfrac{q_{w}}{\rho_{\infty} U_{\infty} c_{p} (T_{0} \, - \, T_{w})}, 
\end{align}
where $q_{w}$ is the heat flux at the surface, $c_{p}$ is the specific heat capacity, and $T_{0}$ is the stagnation temperature. In experiments, the Stanton number can be inferred from TSP and infrared thermography measurements}.  Figure~\ref{fig:base}(c) compares experimental values of $St$ to those predicted by our 2D simulations at different grid resolutions. We see that the computed flow captures the heat flux trends correctly except near the separation and the post-reattachment regions. In experiments, these regions display significant spanwise variation in $St$ and they are marked by the grey band in figure~\ref{fig:base}(c).

Since the flow is globally stable with respect to 3D perturbations~\citep{sidharth2018onset}, we conjecture that spanwise variations arise from non-modal amplification of 3D perturbations around the 2D base flow. To verify our hypothesis, we employ global I/O analysis to quantify the amplification of exogenous disturbances and uncover mechanisms that can trigger the early stages of transition in a hypersonic compression ramp flow.

	\vspace*{-1ex}
    \subsection{Frequency response analysis}
    \label{sec.main.IO}
We utilize frequency response analysis to investigate the amplification of infinitesimal upstream perturbations in a hypersonic compression ramp flow. This choice is motivated by the experimental studies~\citep{chuvakhov2017effect,re2017experimental} where variation in the properties of the incoming boundary layer were found to have profound effects on the downstream streaks. By proper selection of the matrix $\B B$ in Eq.~\eqref{eq:SScomp}, we restrict the inputs to the domain prior to separation (i.e., $x/L<0.5$). Furthermore, we choose the perturbation field in the entire domain as the output, $\bphi = \B q$, by setting ${\B C} = {\B I}$. The I/O analysis is conducted on a grid with $412$ cells in the streamwise and $249$ cells in the wall-normal direction (labeled as G3 in figure~\ref{fig:base}(c)). Numerical sponge boundary conditions are applied near the leading edge ($x/L<0.02$) and the outflow ($x/L>1.6$).

\begin{figure}
\centering
\includegraphics[width=0.99\linewidth]{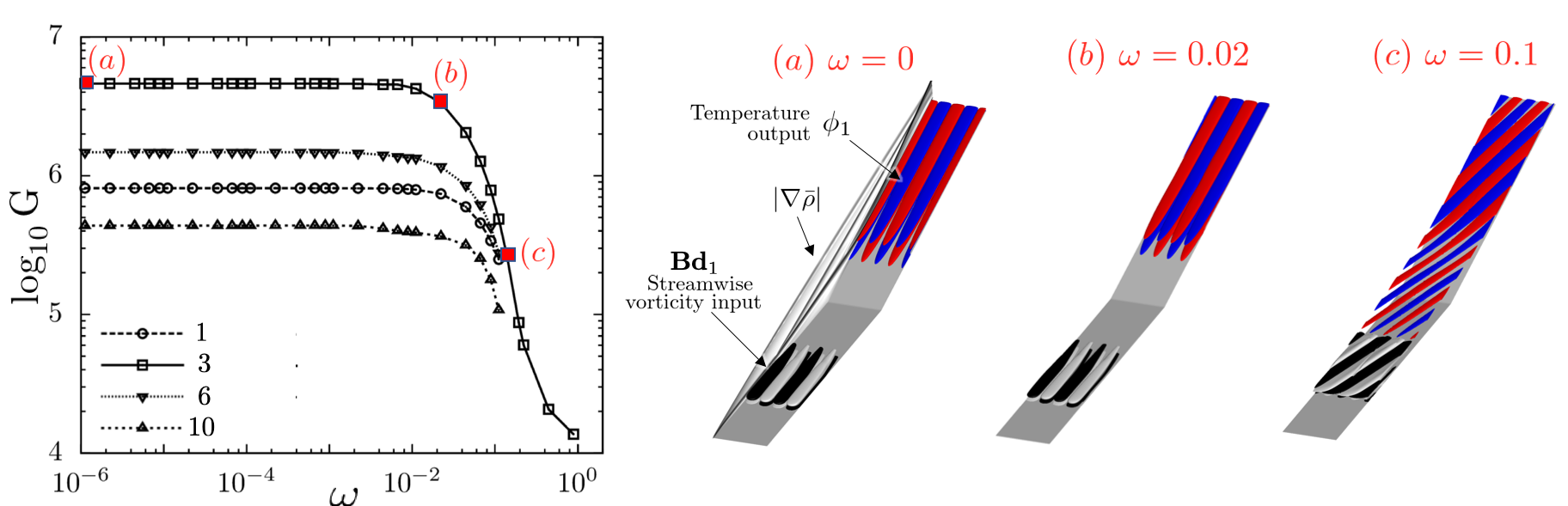}%{unsteady_apr27.png}
\caption{Left: the $\omega$-dependence of the largest induced gain with respect to the compressible energy norm, $\text{G} (\omega) $, for unsteady inputs with the spanwise wavelengths $\lambda_z = \{1, 3, 6, 10 \}$. Right: isosurfaces of streamwise vorticity corresponding to the input $\mathbf{d}_{1}$ and temperature corresponding to the output $\bphi_{1}$ for (a) $\omega = 0$; (b) $\omega =0.02$; and (c) $\omega =0.1$ for $\lambda_{z} = 3$.}
\label{fig:unsteady1}
\end{figure}

\indent The left plot in figure~\ref{fig:unsteady1} shows the input-output amplification $\text{G} (\omega)$, defined in Eq.~\eqref{eq.G}, in a flow with high Reynolds number ($Re_{L} = 3.7\times10^{5} $) for different spanwise wavelengths $\lambda_{z}$. Here, $\lambda_{z} \DefinedAs \lambda^{*}_{z}/\delta_\mathrm{sep}$ and $\omega \DefinedAs \omega^* \delta_\mathrm{sep}/U_{\infty}$ denote the non-dimensional spanwise wavelength and temporal frequency, respectively, $\lambda^*_{z}$ and $\omega^*$ are the corresponding quantities in physical units, whereas $\delta_\mathrm{sep}$ represents the displacement boundary layer thickness at separation. We observe the low-pass feature of the amplification curve: $\text{G}$ achieves its largest value at $\omega = 0$, it decreases slowly for low frequencies, and it experiences a rapid decay after the roll-off frequency ($\omega \approx 0.01$). The visualization of the dominant input-output directions $\mathbf{d}_1$ and $\bphi_{1}$ in figures~\ref{fig:unsteady1}(a) and~\ref{fig:unsteady1}(b) reveals that the flat region of the amplification curve corresponds to incoming streamwise vortical disturbances (as inputs) that generate streak-like downstream perturbations (as outputs). In contrast, figure~\ref{fig:unsteady1}(c) demonstrates that, at high temporal frequency ($\omega = 0.1$), dominant input-output pairs exhibit streamwise periodicity and take the form of oblique waves. It should be noted that the low-pass frequency response features as well as the resulting changes in the response shape (from streaks to oblique waves) were also observed in canonical channel and boundary layer flows~\citep{mj-phd04,ranzarhacjovPRF18}.

\begin{figure}
\centering
\includegraphics[width=0.99\linewidth,trim=1 1 0 0,clip]{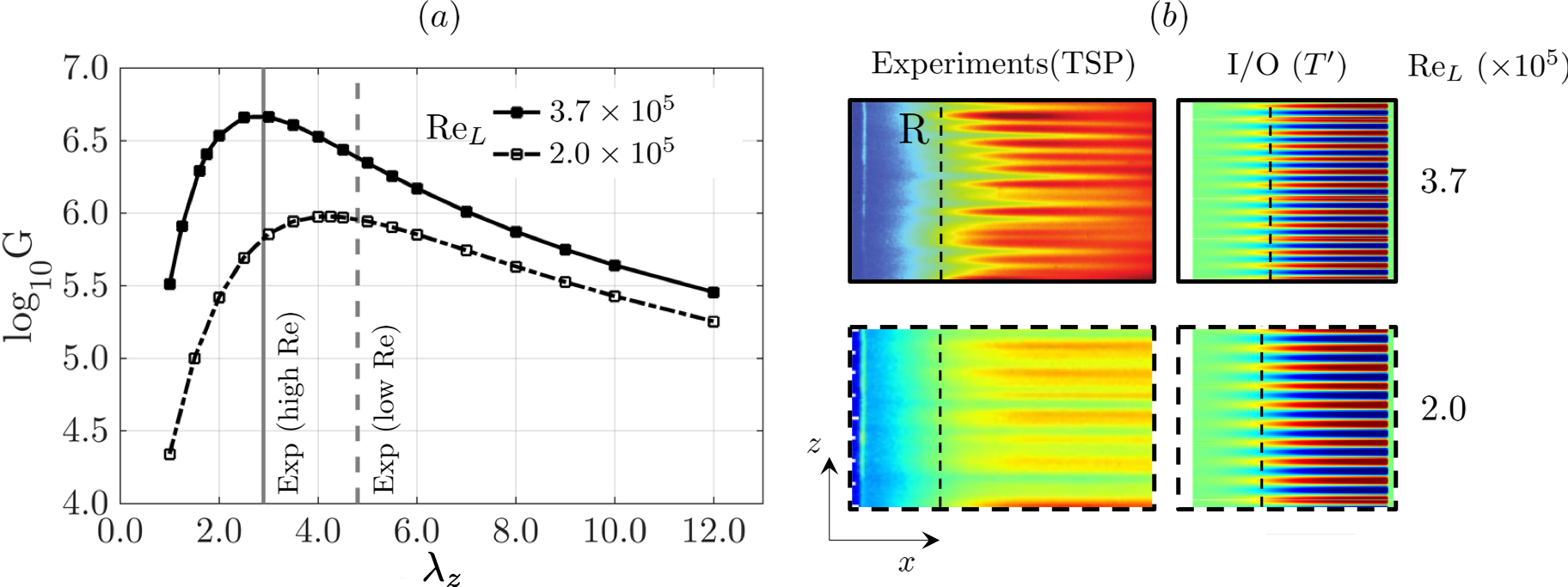}
\caption{(a) The $\lambda_z$-dependence of the amplification map $\text{G}$ for steady inputs (i.e., $\omega=0$); and (b) comparison of experiments and dominant output $\bphi_{1}$ at reattachment. The vertical dashed line in (b) denotes the approximate reattachment line in experiments and 2D simulations.}
\label{fig:Gsteady}
\end{figure}

\indent The impact of the spanwise wavelengths $\lambda_z$ on the amplification $\text{G}$ for steady perturbations (i.e., at $\omega = 0$) is shown in figure~\ref{fig:Gsteady}(a). For both Reynolds numbers, the amplification curve achieves its maximum for a particular value of $\lambda_z$. This indicates that SBLI preferentially amplifies upstream perturbations with a specific spanwise wavelength. %The experimental estimates of $\lambda_z$ resulting from the TSP images in figure~\ref{fig:Gsteady}(b) agree well with the predictions of our I/O analysis.
 The experimental estimates of $\lambda_z$ resulting from the {observed spanwise modulations in the} TSP images in figure~\ref{fig:Gsteady}(b) agree well with the predictions of our I/O analysis. 
Even though the value of $\lambda_z$ at which $\text{G} (0)$ peaks, changes from $\bar{\lambda}_z = 3$ at $Re_L = 3.7\times10^{5}$ to $\bar{\lambda}_z = 4.5$ at $Re_L = 2 \times10^{5}$, the ratio between $\bar{\lambda}_z$ and the displacement boundary layer thickness at reattachment $\delta_R$ remains constant ($\bar{\lambda}_z/\delta_R \approx 1.8$). This value is also consistent with previous studies \citep{inger1977three,navarro2005numerical}.

Our analysis shows that the compression ramp flow strongly amplifies steady upstream disturbances with a preferential spanwise length scale. To understand the effect of disturbances in the recirculation region, we repeat the analysis for $\B B= I$ in Eq.~\eqref{eq:SScomp} and $\omega=0$. We find that the streaks with the same spanwise wavelengths still undergo the largest amplification. This demonstrates that the compression ramp flow is most sensitive to the upstream disturbances, which is consistent with experimental observations.

%\vspace*{0.1cm}
\subsection{Validation of dominant output directions using DNS}
 \label{sec.main.DNS}

We validate the response of the compression-ramp shock/boundary layer interaction to external inputs using 3D DNS of the flow with $Re_{L}=3.7 \times 10^{5}$. The simulations are done in the presence of the dominant input $\mathbf{d}_{1}$ resulting from the I/O analysis at $\omega = 0$ and $\lambda_{z}=3$. The amplitude of the input is fixed at $0.01 \%$ of the of the corresponding free-stream values given in Table~\ref{tab:fs}. We employ Crank-Nicolson implicit time marching scheme and low-dissipation second-order fluxes for spatial discretization. To accurately capture the evolution of the 3D perturbations, the CFL number is set to 10 (i.e., the time step in physical units is around $10 \,\mathrm{ns}$). We employ periodic boundary conditions in the spanwise direction, use 32 grid points for resolving the spanwise wavelength of $3$, and find our results to be independent of the spanwise width of the domain.

\begin{figure}
\centering
\includegraphics[width=0.95\linewidth]{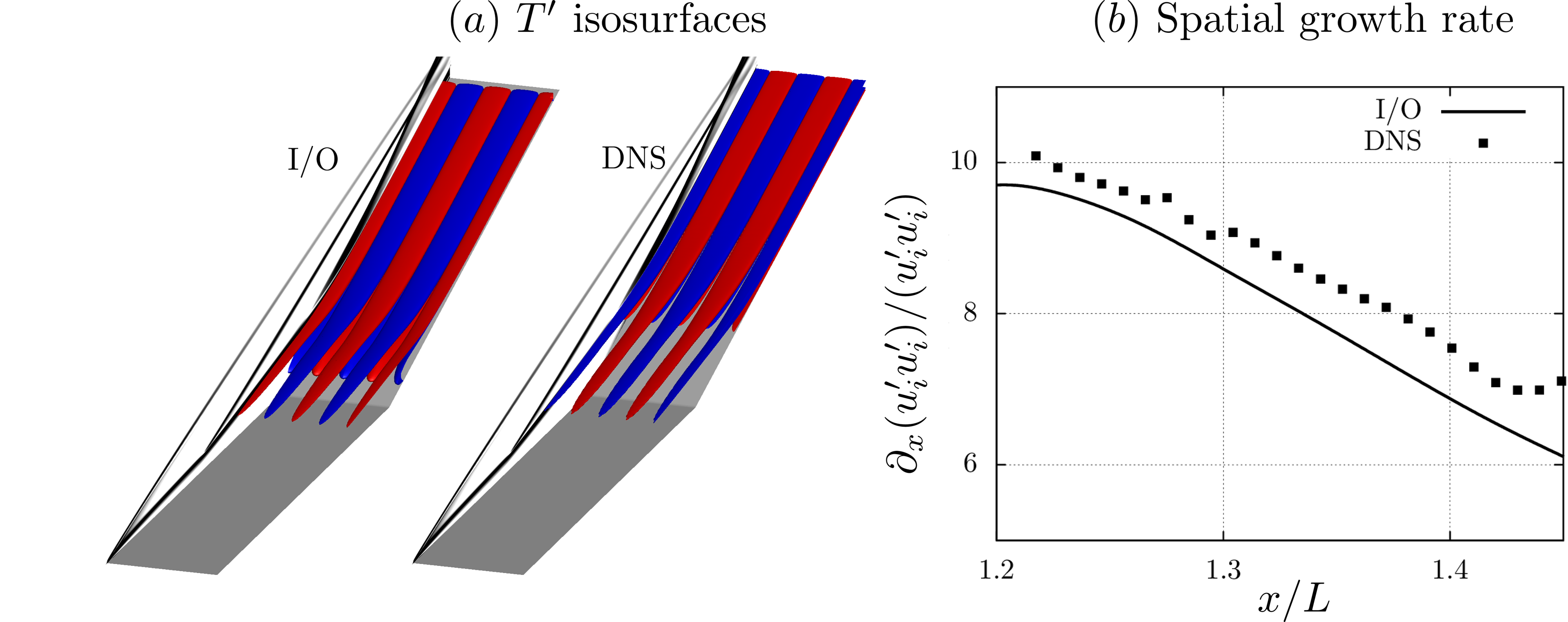}
\caption{Comparison of (a) the temperature perturbations; and (b) the streamwise growth rate of streamwise perturbation energy corresponding to DNS and the dominant output $\bphi_{1}$ resulting from the I/O analysis at $\lambda_{z}=3$. The reattachment location is at $x/L = 1.34$.}
\label{fig:IOvsDNS}
\end{figure}

\indent Figure~\ref{fig:IOvsDNS} demonstrates qualitative similarity between (i) the spatial structure of the temperature perturbations $T'$; and (ii) the spatial growth rate of the perturbation specific kinetic energy resulting from the I/O analysis and the DNS. Even though the DNS results validate the predictions of our analysis, other mechanisms for streak formation are possible. For example, it is well known that unsteady oblique modes can interact nonlinearly to produce streaks in canonical flows~\citep{schmid1992new, berlin1994spatial,fasel1993direct,sandham1995direct,chang1994oblique}.  However, even when we conduct simulations using a pair of unsteady oblique inputs (shown in figure~5(c) and with the same amplitude as the previous steady inputs), the dominant responses in DNS are still given by the steady streaky outputs. Since both I/O analysis and DNS identify steady streaks as the robust flow features, the amplification of infinitesimal upstream perturbations may play an important role in the formation of the streaks in realistic flow configurations. Thus, in what follows, we investigate the spatial structure of the dominant steady responses resulting from the I/O analysis in an attempt to uncover physical mechanisms responsible for the streak formation in compression ramp flow.

	\vspace*{-2ex}
\subsection{Spatial structure of the most amplified perturbations}\label{sec.main.structure}
 
\begin{figure}
\centering
\includegraphics[width=0.9\linewidth,trim=0 0 0 0,clip]{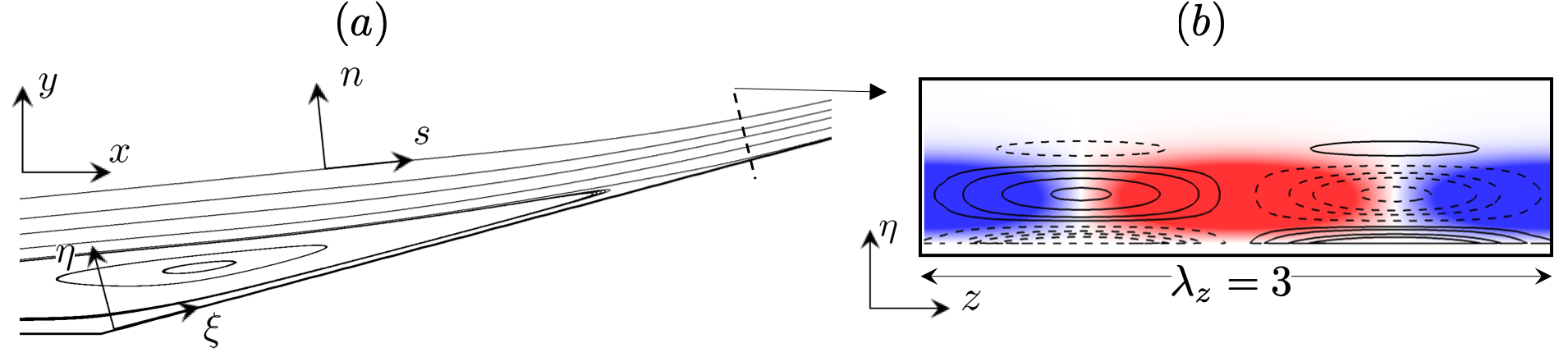}
\caption{(a) Schematic of the different coordinate systems for analyzing the perturbation evolution; and (b) color plots of streamwise velocity $u'_{s}$ and contour lines of streamwise vorticity $\omega'_{s}$ at streamwise location $x/L=1.4$ (post-reattachment) corresponding to the dominant output $\bphi_{1}$ at $\lambda_{z}=3$. Solid lines denote positive values and dashed lines denote negative values of $\omega'_{s}$.}
\label{fig:gortler}
\end{figure}

In order to gain insight into the spatial structure of the most amplified perturbations, we examine the velocity and vorticity components ($u'_s,\omega'_s$) of the dominant output $\bphi_1$ along the coordinate system associated with the base flow streamlines~\citep{bradshaw1973effects}. In~figure~\ref{fig:gortler}, we also show the wall-aligned coordinate system, where $\xi$ and $\eta$ denote the directions parallel and normal to the wall, respectively. In the flow with $Re_{L} = 3.7 \times 10^5$, figure~\ref{fig:gortler}(b) illustrates the output components corresponding to $\lambda_z = 3$ near reattachment in the $(\eta,z)$ plane. We note that the most amplified perturbations are given by alternating regions of high and low velocities with counter rotating vortices between them and that $u'_{s}$ and $\omega'_{s}$ are $90^{\circ}$ out of phase in the spanwise direction. 

\begin{figure}
\centering
\includegraphics[width=0.95\linewidth,trim=0 0 0 0,clip]{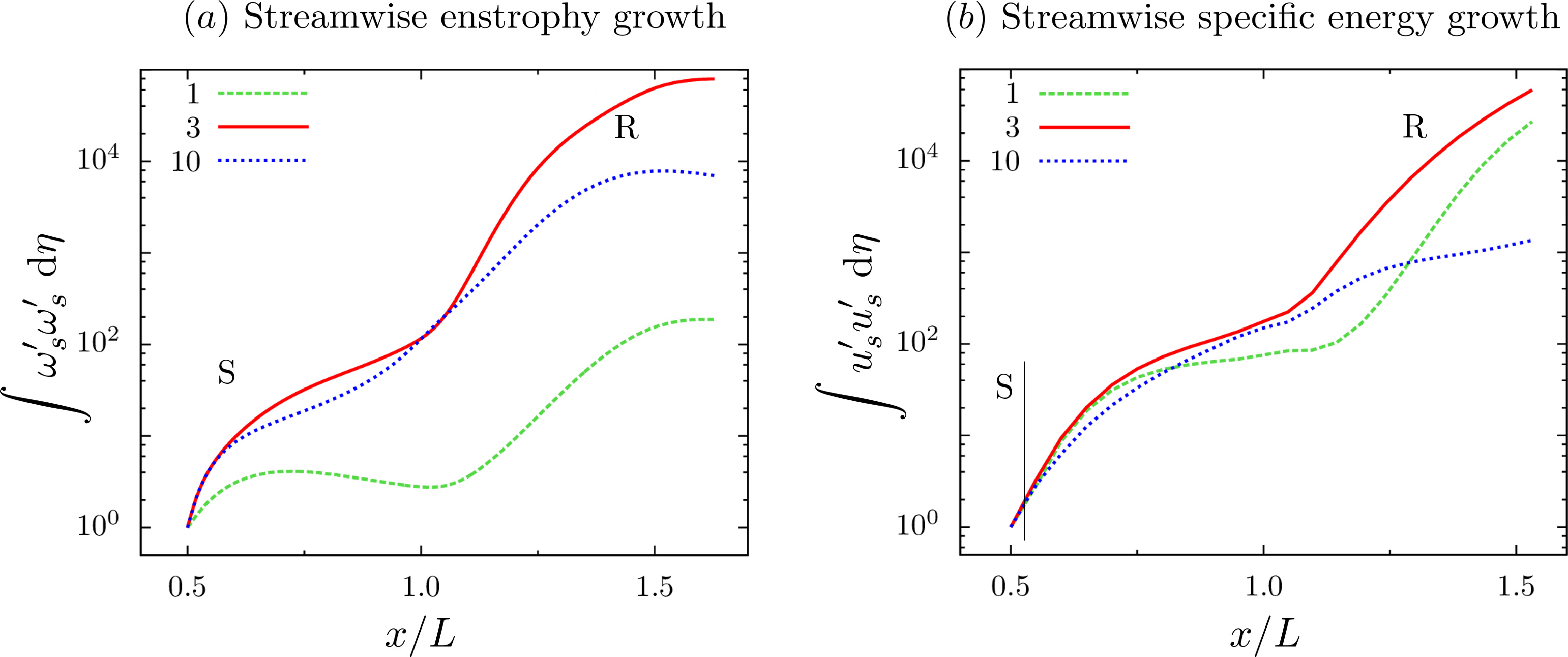}
\caption{Spatial evolution of the wall-normal integral of (a) streamwise enstrophy; and (b) streamwise specific energy of the dominant output $\bphi_{1}$ for $\lambda_{z}=1,3,10$. The lines $S$ and $R$ denote the separation and reattachment points in the 2D base flow. The values are normalized using the respective wall-normal integrals at $x/L=0.5$.}
\label{fig:evol}
\end{figure}

\begin{figure}
\centering
\includegraphics[width=0.99\linewidth]{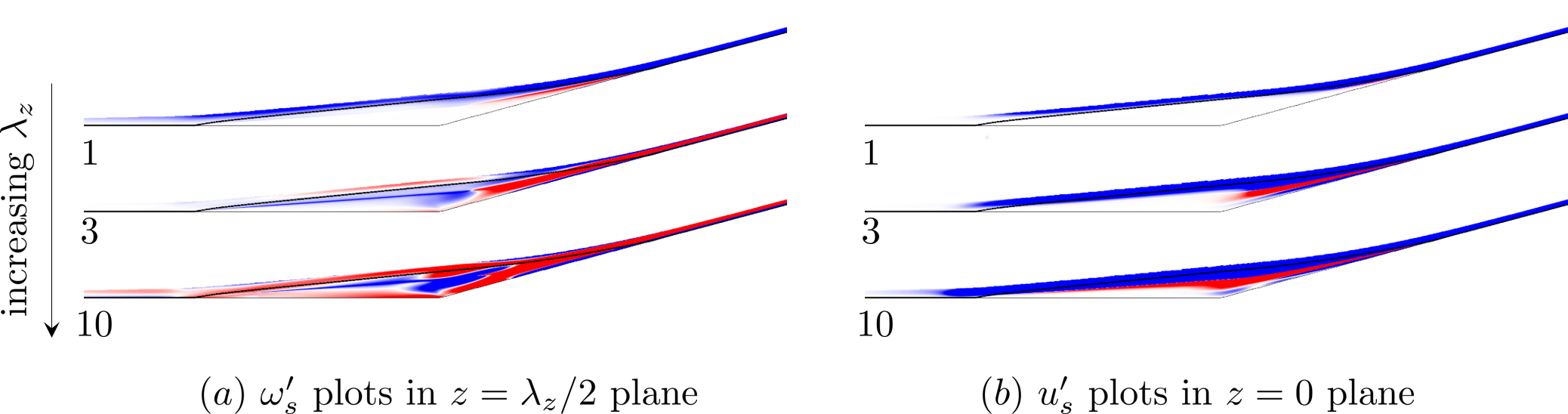}
\caption{Streamwise (a) vorticity; and (b) velocity perturbations corresponding to the dominant output $\bphi_{1}$ in the ($x,y$) plane. The bold black line denotes the separation streamline. Note that $\omega'_s$ is $90^\circ$ out of phase with respect to $u'_s$ in the spanwise direction.}
\label{fig:flood}
\end{figure}

\indent To quantify the spatial evolution of flow perturbations, we compute the wall-normal integrals of the streamwise enstrophy ($\omega'_{s}\omega'_{s}$) and the streamwise specific energy ($u'_su'_s$) as a function of $x$ for three different values of $\lambda_{z}$ ($1$, $3$, and $10$). In the flow with $Re_{L} = 3.7 \times 10^5$, these respectively identify the outputs with small, dominant, and large spanwise wavelengths. To ensure that the perturbations in the separated shear layer are captured, the wall-normal integral is computed for $\eta \in [0, 5\delta_{\mathrm{sep}} ]$ where $\delta_{\mathrm{sep}}$ is the displacement boundary layer thickness at separation. For $\lambda_{z} = 1$, figure~\ref{fig:evol} illustrates that both the streamwise enstrophy and specific energy saturate within the bubble followed by a large amplification near the reattachment~R. In contrast, for $\lambda_{z} = 10$ the perturbations grow steadily in the bubble followed by a weaker amplification near reattachment. For $\lambda_{z} = 3$, the flow perturbations experience significant amplification in both the separated zone (prior to the corner, $x/L \approx 1$) and in the reattachment region. These amplification trends are further illustrated in figure~\ref{fig:flood} which visualizes $\omega'_s$ and $u'_{s}$ in the ($x,y$) plane. We see that both $\omega'_s$ and $u'_{s}$ have footprints inside the recirculation zone which demonstrates that they do not solely reside in the reattaching shear layer. The strength of the perturbations in the recirculation zone increases with increase in $\lambda_{z}$. Therefore, the spanwise wavelength where largest amplification occurs is associated with vortical perturbations with significant contribution from both the separation zone (i.e., the separation bubble and the shear layer) and the reattaching boundary layer. In what follows, we refer to these
steady perturbations as `reattachment streaks'.

	%\vspace*{0.2cm}
\subsection{I/O analysis without separation bubble perturbations}
 \label{sec.main.bubnbub}
\begin{figure}
\centering
\includegraphics[width=1.0\linewidth]{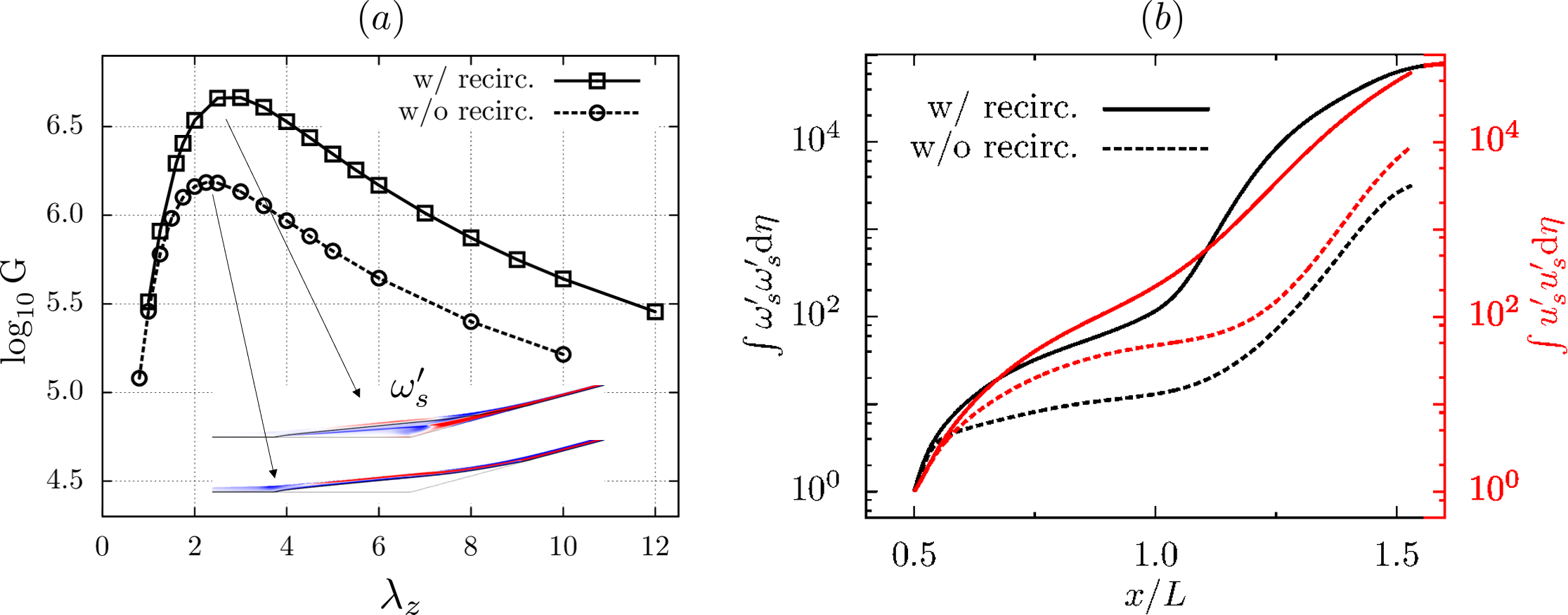}
\caption{(a) Amplification map $\text{G}$ for steady inputs along with insets of streamwise vorticity $\omega'_{s}$; and (b) comparison of wall normal integrals of the streamwise enstrophy ($\omega'_{s}\omega'_{s}$) and specific kinetic energy ($u'_{s}u'_{s}$) in the presence and absence of perturbation dynamics in the recirculation bubble for $\lambda_{z}$ that yields the largest amplification. The values in (b) are normalized using the respective wall-normal integrals at $x/L=0.5$.}
\label{fig:nobub}
\end{figure}

\indent To confirm the role of separation bubble in the amplification of flow perturbations, we carry out the I/O analysis by excluding the perturbation dynamics in the bubble. For $Re_{L} = 3.7\times 10^{5}$, we introduce the 2D base flow separation streamline as an artificial boundary, ensuring that the incoming streamwise vortices travel parallel to the surface of this inviscid boundary with the curvature properties of the separation streamline. Thus, no perturbations enter the recirculation zone, and all of them are equal to zero inside this region, as seen in figure~\ref{fig:nobub}(a) for $\omega'_s$.

\indent The amplification map in figure~\ref{fig:nobub}(a) shows that eliminating the role of the bubble perturbations reduces the largest amplification five times. Also, the spanwise wavelength that corresponds to the largest gain decreases from $\lambda_z = 3.0$ to $\lambda_z = 2.25$. Figure~\ref{fig:nobub}(b) shows that the wall-normal integrals of the streamwise enstrophy ($\omega'_{s}\omega'_{s}$) and streamwise specific energy ($u'_{s}u'_{s}$) are also reduced in the absence of the perturbation dynamics within the bubble. The perturbation growth in the present case clearly saturates until the reattachment region, beyond which it follows the same trend as in the original I/O output fields. We note that, for small values of $\lambda_{z}$, the I/O analysis of the shock/boundary layer interaction reveals that perturbations experience significant amplification near reattachment without any contribution from the recirculation bubble. This is consistent with figure~\ref{fig:nobub}(a), which shows almost identical gains for small values of $\lambda_{z}$.

%\vspace*{0.15cm}
\vspace*{-3ex}
\section{Amplification of steady reattachment streaks: physical mechanism}
\label{sec.physics}

\indent As demonstrated in the previous section, the hypersonic flow over a compression ramp selectively amplifies small upstream perturbations of a specific spanwise wavelength. The largest amplification is associated with steady perturbations where different regions of the 2D base flow contribute to the growth of 3D reattachment streaks. To characterize the streak amplification, we examine the equation that governs the evolution of Chu's compressible energy $\Ece$~\citep{chu1965energy,hanifi1996transient} of the perturbations resulting from the I/O analysis. As shown in~\cite{sidharth2018onset}, this equation is given~by
\begin{align}
    \label{eq:chu_prod}
    \dfrac{\mrd \Ece}{\mrd t}    
    \, + \, 
    \mathcal{T} 
    \; = \; 
    \mathcal{P} \, + \,  \mathcal{S} \, + \,  \mathcal{V} \, + \,  \mathcal{F},
\end{align}
where $\mathcal{T}$, $\mathcal{S}$, $\mathcal{P}$, and $\mathcal{V}$ respectively determine transport, source, production, and viscous terms~\cite[Eqs.\ ($16$) and ($17$)]{sidharth2018onset}. On the other hand, $\mathcal{F}$ accounts for the work done by external disturbances; see Appendix~\ref{appendixC}. The transport term $\mathcal{T}$ is responsible for advection of perturbations by the base flow velocity; the source term $\mathcal{S}$ corresponds to the perturbation component of the inviscid material derivative; the production term $\mathcal{P}$ quantifies interactions of perturbations with the mean flow gradients and is, in general, sign-indefinite; and the viscous term $\mathcal{V}$ determines dissipation of Chu's compressible energy by viscous stresses.

\begin{figure}
\centering
\includegraphics[width=0.75\linewidth]{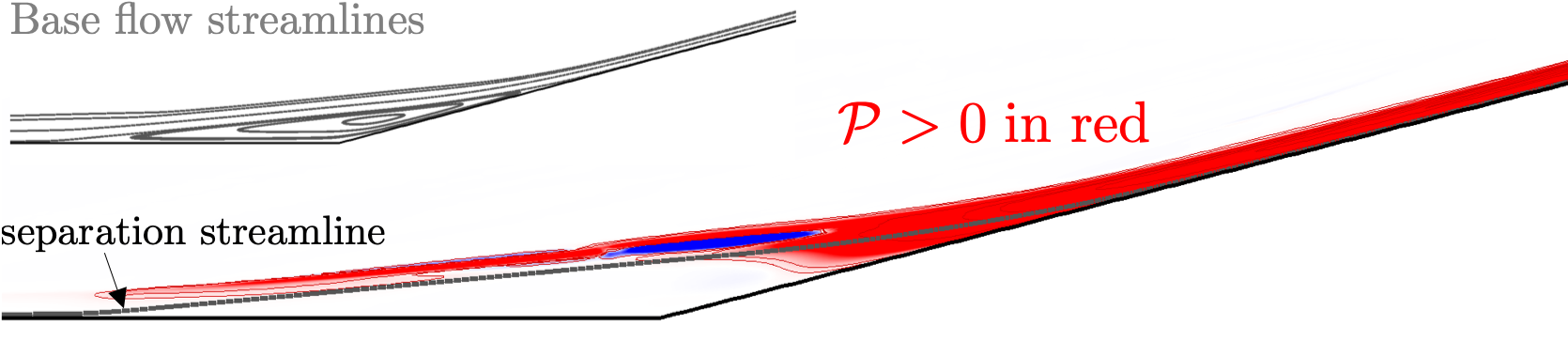}%fig1_enlarged.png}
\caption{Distribution of the production term $\mathcal{P}$ in Eq.~\eqref{eq:chu_prod} for steady perturbations with $\lambda_{z}=3$. The regions marked in red and blue correspond to positive and negative values of $\mathcal{P}$.}
\label{fig:prod_chu}
\end{figure}

For incompressible flows, the divergence-free property of velocity field can be utilized to simplify the terms in Eq.~\eqref{eq:chu_prod}; e.g., see~\citet{sipp2013characterization} for the analysis of energy amplification in incompressible spatially-developing boundary layer. Here, we evaluate different terms in Eq.~\eqref{eq:chu_prod} for the most amplified steady perturbations with $\lambda_{z}=3$. The viscous terms are dissipative and, thus, negative throughout the domain and the terms $\mathcal{S}$ and $\mathcal{F}$ are found to be negligible. As illustrated in figure~\ref{fig:prod_chu}, the production term $\mathcal{P}$ associated with steady perturbations is active in the spatial locations near and after reattachment. Further analysis reveals that the dominant positive contribution to $\mathcal{P}$ comes from the momentum transport equation.  

To investigate the physical mechanisms responsible for the amplification of 3D reattachment streaks we analyze the dominant terms in the linearized inviscid transport equations. In particular, we examine the spatial development of the streamwise vorticity, velocity, and temperature perturbations and identify amplification mechanisms that result from the interactions of flow perturbations with base flow gradients.

	\vspace*{-2ex}
\subsection{Inviscid transport of streamwise vorticity}\label{sec.physics.oms}

\indent We consider transport of flow fluctuations in the $(s,n,z)$ coordinate system which is locally aligned with the streamlines of the base flow $\big( \bar{u}_{s},0,0 \big)$. This coordinate system allows us to quantify relative contributions to the energy amplification of base flow gradients, the flow curvature in the presence of flow separation, and the baroclinic effects. Similar  framework has been utilized to evaluate the effect of longitudinal streamline curvature on Reynolds stresses in turbulent boundary layer and shear layer flows~\citep{finnigan1983streamline,kansa2002local,patel1997longitudinal,maurizi1997method,richmond1986equations}.

\indent As shown in Appendix~\ref{appendixB}, the inviscid transport of steady streamwise vorticity perturbation $\omega'_s$ can be written as 
\begin{align}
     \bar u_s \partial_s \omega'_s  
     \; \approx  \; 
      \frac{ \partial_n  \bar \rho}{\bar \rho^2 } \, \mri\beta p'
     \; - \;
     \partial_n \bar u_s
     \,
     \partial_s w'  
     \; - \; 
     \dfrac{2 \bar u_{s}}{\mathcal{R}}  \, { \mri\beta u'_{s} }, \label{eq:vortT} 
      \end{align}
where $\mathcal{R}$ is the local radius of curvature, $\big (u'_{s},u'_{n},w' \big)$ are the velocity fluctuations, and $\omega'_s \DefinedAs \partial_n w' - \mri \beta u_n'$ where $\beta \DefinedAs 2\pi/\lambda_{z}$. The left-hand-side of Eq.~\eqref{eq:vortT} determines streamwise advection of $\omega_s'$ by the base flow $\bar{u}_{s}$. On the other hand, the three terms on the right-hand-side  lead to production ($\mathcal{P}$) of $\omega'_{s}$ and they account for:
\begin{itemize}
	\item baroclinic effect, which arises from misalignment of pressure and density gradients and accounts for differential acceleration caused by variable inertia~\citep{candler2014baroclinic,sidharth2018subgrid};
  \item vortex tilting, which redistributes vorticity perturbations from the stream-normal direction $n$ to the streamwise direction $s$; and
  \item centrifugal effect, which originates from the curvature $1/\mathcal{R}$ in the coordinate system associated with the base flow streamlines. 
\end{itemize}

\begin{figure}
\centering
\includegraphics[width=0.999\linewidth,trim=0 0 0 0,clip]{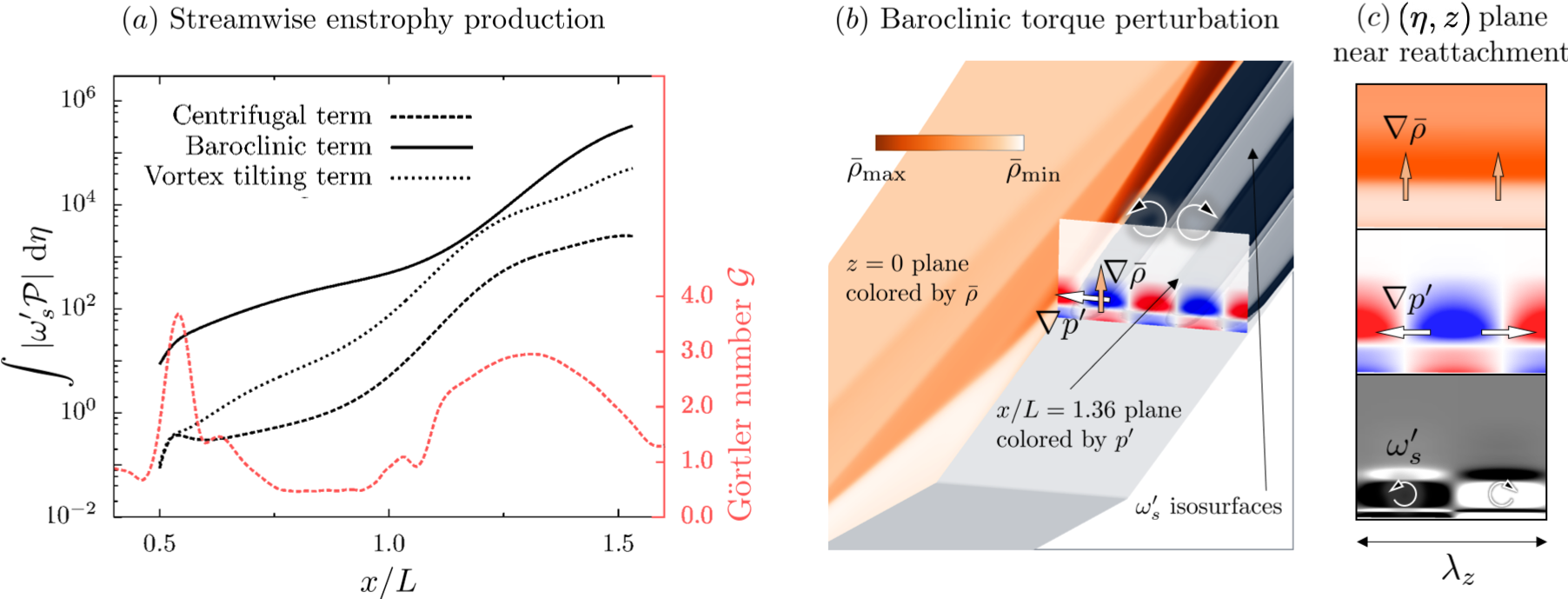}
\caption{(a) Comparison of the magnitude of the production terms $\mathcal{P}$ in Eq.~\eqref{eq:vortT1} (where the perturbations are normalized using the value of $\int \omega'_s \omega'_s \mrd \eta$ at $x/L = 0.5$) along with the streamwise variation of the G\"ortler number $\mathcal{G}$; (b) illustration of the baroclinic term $\nabla \bar \rho \times \nabla p'$ for the dominant output with $\lambda_z=3$; and (c) corresponding quantities near the reattachment plane at $x/L=1.36$.}
\label{fig:ombud}
\end{figure}

	\vspace*{0.15cm}
	
	\indent 
Multiplication of Eq.~\eqref{eq:vortT} with $\omega_s'$ and integration over $\eta$ yield the equation that can be used to evaluate the spatial transport of streamwise enstrophy ($\omega'_{s}\omega'_{s}$) and assess the relative contribution of different physical effects,
	\begin{equation}
	\begin{array}{rcl}	
	\dfrac{1}{2} \, 
	\displaystyle{\int_{0}^{\eta_0} }
	\bar u_s \partial_s (\omega'_{s}\omega'_{s}) \, \mrd \eta
       	& \!\!\! \approx  \!\!\! &
	\displaystyle{\int_{0}^{\eta_0} }
       \frac{ \partial_n  \bar \rho}{\bar \rho^2 } \, \mri\beta p'
       \omega'_s
       \,
       \mrd \eta
       \; - \; 
       \displaystyle{\int_{0}^{\eta_0} }
       \partial_n \bar u_s 
       ( \omega'_s \partial_s w' )
       \, 
       \mrd \eta
       ~ -
       \\[0.45cm]
       & \!\!\! \!\!\! &
      \displaystyle{\int_{0}^{\eta_0} }
       \dfrac{2 \bar u_{s}}{\mathcal{R}}  \, { \mri\beta u'_{s} \omega'_{s}} 
       \,
       \mrd \eta,
     	\end{array}
	 \label{eq:vortT1}
	\end{equation}
where $\eta_0 = 5 \delta_{\mathrm{sep}}$. Figure~\ref{fig:ombud}(a) compares the $x$-dependence of the absolute values of the terms on the right-hand-side of Eq.~\eqref{eq:vortT1} for dominant output perturbations resulting from the I/O analysis with $\lambda_{z}=3$. The dominant contribution arises from the baroclinic effect, with spanwise variations in $p'$ and stream-normal variations in $\bar \rho$ representing the prime sources of the baroclinic torque perturbations. In contrast to incompressible flows, the baroclinic term is particularly important in cold-wall hypersonic boundary layers and it is the sole contributor to $\omega'_s$ in the bulk of the separation zone. 

\indent Previous experimental~\citep{chuvakhov2017effect,roghelia2017experimental} and numerical~\citep{navarro2005numerical} SBLI studies have attributed the development of streamwise streaks to the centrifugal effects. The G\"ortler number, $\mathcal{G} = \sqrt{L/ ( \mathcal{R}\epsilon )}$, where $\epsilon \DefinedAs \bar{u}_{\eta}/\bar{u}_{\xi}$ (see Appendix~\ref{appendixA}), quantifies the effect of local flow curvature and figure~\ref{fig:ombud}(a) shows that contribution of the centrifugal terms to the reattachment streaks increases in the regions of high $\mathcal{G}$ (i.e., near separation and reattachment points). Relative to baroclinic and vortex tilting terms, the centrifugal effects appear to play a minor role in the spatial amplification of reattachment streaks. Since the largest contribution comes from the baroclinic term, our analysis of the spatial transport of the most amplified output perturbations demonstrates that baroclinic effects (rather than centrifugal effects) trigger reattachment streaks in hypersonic compression ramp flows. 

\indent We illustrate the linear baroclinic mechanism in figure~\ref{fig:ombud}(b) by showing three quantities: (i) the base flow density $\bar \rho$ in the ($x,y$) plane using an orange colormap; (ii) the spanwise gradient of the pressure perturbations $p'$ in the $(y,z)$ plane near reattachment using the red-white-blue colormap; and (iii) the isosurfaces of streamwise vorticity $\omega'_s$ using a grey-black colormap. Since the linearized baroclinic torque that is active in the steady response is associated with $\nabla \bar \rho \times \nabla p'$, we focus on examining the gradients of $\bar{\rho}$ and $p'$ shown in figure~\ref{fig:ombud}(c). Near reattachment, the density gradient is aligned with the wall-normal direction $\eta$. This is because the $\bar{\rho}$ colormap becomes darker as we move away from the wall in the direction of increasing $\eta$. At the same $x$ location, the gradient of $p'$ is orthogonal to the ($x,y$) plane; it achieves its largest value midway between the blue and the red lobes and it points in the direction from the center of the blue to the center of red lobes. As illustrated in figures~\ref{fig:ombud}(b) and (c), the resulting linearized baroclinic torque $\nabla \bar \rho \times \nabla p'$ aligns with the streamwise vorticity $\omega'_{s}$, thereby leading to its production. 

	\vspace*{-2ex}
\subsection{Inviscid transport of streamwise velocity}\label{sec.physics.us}
 
Recent experimental~\citep{mustafa2019amplification} and numerical studies~\citep{sandham2014transitional,dwivedi2017optimal} demonstrated that streamwise velocity perturbations contribute most to the kinetic energy. In Appendix~\ref{appendixA}, we confirm this observation using relative scaling of the perturbation quantities. The spatial transport of streamwise velocity $u'_s$  is governed by, 
\begin{align}
        \bar u_s \, \partial_s u'_s  
     \; \approx \; 
     - \partial_s \bar u_s \, u'_s
     \; - \; 
     \partial_n \bar u_s \, u'_n 
     \; - \; 
    \dfrac{\bar u_s }{\mathcal{R}} \, u'_n  
     \; - \; 
     \dfrac{1}{\bar \rho}\,\partial_s p', \label{eq:us}
\end{align}
where the term on the left-hand-side quantifies the transport of $u_s'$ by the base flow $\bar{u}_s$. The first two terms on the right-hand-side account for the production ($\mathcal{P}$) of perturbations by the base flow gradients. In particular, the first term is responsible for the growth of $u_s'$ because of the streamwise deceleration of the base flow (where $\partial_{s} \bar u_{s}< 0$), the second term originates from the base flow shear and it accounts for the lift-up mechanism~\citep{landahl1980note}, and the additional terms account for the centrifugal effects and the influence of pressure gradient. 

\indent Multiplication of Eq.~\eqref{eq:us} with $u_s'$ and integration over $\eta$ yield the equation that can be used to evaluate the spatial transport of streamwise specific energy ($u'_{s}u'_{s}$) and assess the relative contribution of different physical effects,
	\begin{equation}
	\begin{array}{rcl}	
	\dfrac{1}{2} \, 
	\displaystyle{\int_{0}^{\eta_0} }
	\bar u_s \, \partial_s (u'_{s}u'_{s}) \, \mrd \eta
       	& \!\!\! \approx  \!\!\! &
	- \displaystyle{\int_{0}^{\eta_0} }
	\partial_s \bar u_s\, (u'_s u'_{s})\, \mrd \eta  
       \; - \; 
       \displaystyle{\int_{0}^{\eta_0} }
       \partial_n \bar u_s \, (u'_n u'_s) 
        \, 
       \mrd \eta
       ~ -
       \\[0.45cm]
       & \!\!\! \!\!\! &
      \phantom{-} \displaystyle{\int_{0}^{\eta_0} }
       \dfrac{\bar u_{s}}{\mathcal{R}}  \, (u'_{n} u'_{s})
       \,
       \mrd \eta \;-\; \displaystyle{\int_{0}^{\eta_0} }
       \dfrac{1}{\bar \rho} \, (\partial_{s}p' \, u'_{s}) 
        \, 
       \mrd \eta  
       ,
     	\end{array}
	 \label{eq:us1}
	\end{equation}

\noindent where $\eta_{0}=5\delta_{\mathrm{sep}}$. The centrifugal effects are found to be negligible and the pressure gradient reduces  growth of specific kinetic energy. Thus, to quantify the spatial amplification of $u'_{s}$ it is essential to examine the role of the production terms (i.e., the first two terms on the right-hand-side of Eq.~\eqref{eq:us1}).

\begin{figure}
\centering
\includegraphics[width=0.999\linewidth,trim=0 0 0 0,clip]{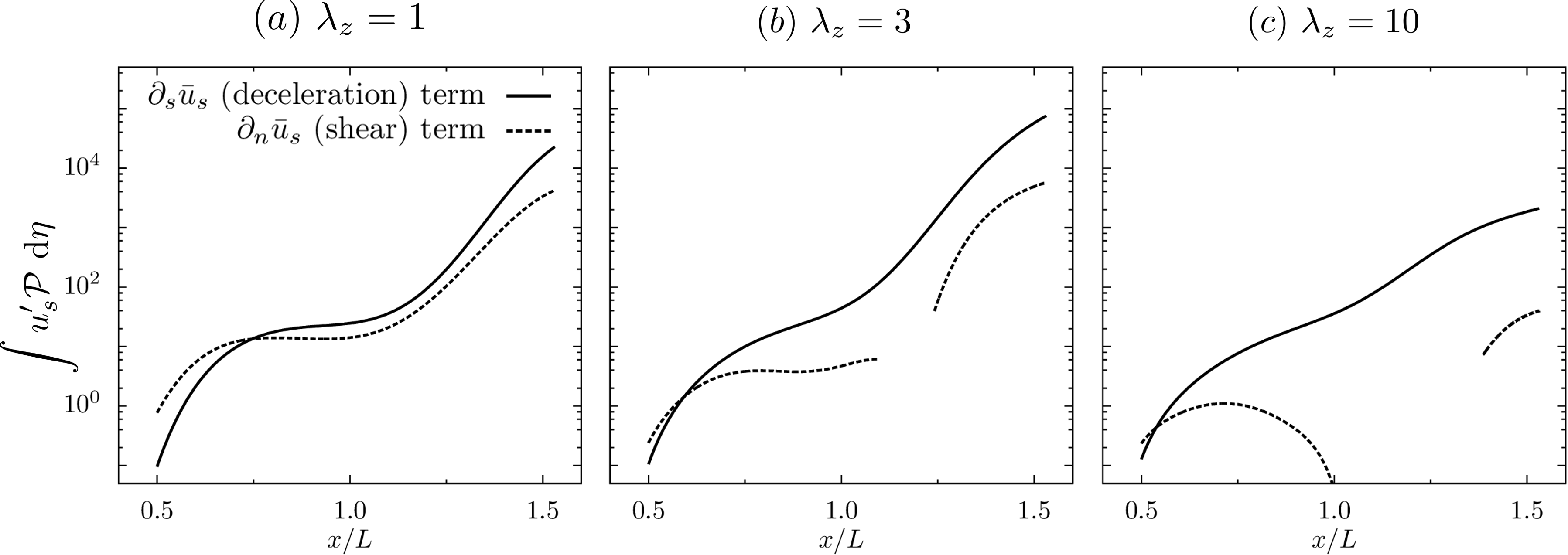}%{usb_apr8.png}
\caption{Contribution of the production term $\mathcal{P}$ in Eq.~\eqref{eq:us1} for the output $\bphi_{1}$ with small, dominant, and large value of $\lambda_z$. For each spanwise wavelength, the terms are normalized using the wall-normal integrals of the streamwise specific kinetic energy (at $x/L = 0.5$). The negative values of the shear term are not plotted.}
\label{fig:usb}
\end{figure}

\indent The contribution of the production terms to the streamwise specific kinetic energy is illustrated in figure~\ref{fig:usb} for three different values of $\lambda_{z}$. In all three cases, streamwise deceleration term dominates the production of $u'_{s}$ and it peaks for the spanwise wavelength $\lambda_{z}=3$. The lift-up effect introduces a large positive contribution for the perturbations with small values of $\lambda_{z}$. These perturbations are almost absent in the recirculation zone and the contribution from this effect is dominant prior to separation. For larger values of $\lambda_{z}$, the contribution of lift-up mechanism decreases after separation and becomes negative over a significant region within the separation zone. This explains the reduced amplification at reattachment for large spanwise wavelength observed in figure~\ref{fig:evol}(b).

	\vspace*{-2ex}
\subsection{Inviscid transport of temperature perturbations}

\indent To understand the formation of heat streaks near reattachment, we consider the spatial amplification of temperature perturbations $T'$ as they are transported by the base flow. For the most amplified output perturbations, we retain the terms with significant contribution to the inviscid transport equation for $T'^2$,
\begin{align}
{\bar u_s} \, \partial_s (T'^2/2)
	\; &\approx \; 
	-  \partial_s \bar T \, (u'_s T')
	\; - \; 
	\partial_n \bar T \, (u'_n T')
	\; - \; 
	(\gamma-1) \, ( \nabla \cdot \bar{\bm{u}} ) \, T'^2.
	\label{eq:T1}
	\end{align}
It turns out that $\partial_n w'$ does not have significant contribution to the production of $T'^2$ for $\lambda_{z} = 3$ and, since $\omega'_s = \partial_n w' - \mri \beta u'_n$, we have $u'_n \approx -\omega'_s/(\mri \beta)$. Thus, for the most amplified steady 3D output perturbations, we can approximate Eq.~\eqref{eq:T1} as
\begin{align}
{\bar u_s} \, \partial_s (T'^2/2)
	\; &\approx \; 
	-  \partial_s \bar T \, (u'_s T')
	\; + \; 
	\partial_n \bar T \,\dfrac{(\omega'_s T')}{\mri \beta} 
	\; - \; 
	(\gamma-1) \, ( \nabla \cdot \bar{\bm{u}} ) \, T'^2. 
	\label{eq:T2}
\end{align}
\noindent The first term on the right-hand-side is associated with the temperature perturbation flux in the streamwise direction
and the streamwise gradient $\partial_s \bar T$ near reattachment is responsible for the production of the temperature fluctuations. The second term accounts for the transport of $T'$ by the streamwise vorticity $\omega'_s$ across the wall-normal thermal base flow gradients in the boundary layer. Therefore, both $u'_s$ and $\omega'_s$ contribute to production of temperature fluctuations at reattachment. The third term quantifies the base flow dilatation in the reattachment shock where $\nabla \cdot \bar{\bm{u}}$ takes large negative values. All of these three physical effects significantly contribute to the amplification of $T'$ near reattachment. 

	%\vspace*{-0.5ex}
	% \vspace*{1ex}
\begin{remark}
{\em Our analysis of inviscid transport equations uncovers physical mechanisms responsible for the amplification of steady reattachment streaks. We showed that streamwise deceleration contributes to the amplification of $u'_{s}$ and  that the baroclinic effects are responsible for the amplification of $\omega'_{s}$. Furthermore, the appearance of the temperature streaks near reattachment is triggered by the growth of both $u'_{s}$ and $\omega_{s}'$ as well as by the amplification of upstream temperature perturbations by base flow dilatation $\nabla\cdot \bar{\bm{u}}$ that originates from the reattachment shock. The spanwise scale selection can be attributed to the presence of flow perturbations in the separation bubble and in the reattaching shear layer. As demonstrated in \S~\ref{sec.main.structure}, weak amplification for small spanwise wavelengths arises from the absence of perturbation dynamics within the bubble. In contrast, for large values of $\lambda_{z}$, amplification is weak because the perturbations in the bubble destructively interfere with the perturbations in the separated shear layer; see \S~\ref{sec.physics.us}.}
	\end{remark}
	
	\vspace*{-2ex}
\begin{remark}
{\em The emergence of reattachment streaks in laminar hypersonic SBLI is typically attributed to centrifugal instability that results from the streamline curvature near reattachment~\citep{navarro2005numerical,chuvakhov2017effect,roghelia2017experimental,re2017experimental}. In contrast, our analysis demonstrates the importance of baroclinic terms and shows that the centrifugal effects play only a minor role in the spatial amplification of the streaks. In cold wall hypersonic boundary layers, baroclinic torque results from the interactions of upstream pressure perturbations with base flow density gradients which provides a physical mechanism for the emergence of the reattachment streaks.}
    \end{remark}	

	\vspace*{-4ex}
\section{Concluding remarks}
	\label{sec.remarks}

We have employed an input-output analysis to investigate amplification of disturbances in compressible boundary layer flows. Our approach utilizes global linearized dynamics to study the growth of flow perturbations and identify the spatial structure of the dominant response. For supersonic flat plate boundary layers, we have verified that the I/O approach captures both acoustic and vortical spatial growth mechanisms without any {\it a priori} knowledge of the perturbation form. 

In an effort to explain the heat streaks near reattachment, we have also examined the experimentally observed reattachment streaks in shock/boundary layer interaction on Mach 8 flow over $15^\circ$ compression ramp. In spite of global stability, the I/O analysis predicts large amplification of incoming steady streamwise vortical disturbances with a specific spanwise length scale. The dominant output takes the form of steady streamwise streaks near reattachment and we employ DNS to verify robustness of the identified responses. In addition to an agreement with DNS, our predictions of the most amplified spatio-temporal flow structures agree well with two recent experiments. 

We have also uncovered physical mechanisms responsible for amplification of steady reattachment streaks. This was accomplished by evaluating the dominant contribution of the base flow gradients to the production of streamwise velocity, vorticity, and temperature perturbations in the inviscid transport equations. We have demonstrated that streamwise deceleration in the recirculation bubble and the reattaching shear layer is responsible for the amplification of streamwise velocity perturbations and that the baroclinic effects contribute most to the amplification of streamwise vorticity. Furthermore, the appearance of the temperature streaks near reattachment is triggered by the growth of both streamwise velocity and vorticity along with the amplification of upstream temperature perturbations by the reattachment shock.

The emergence of reattachment streaks is typically attributed to G\"ortler-like centrifugal instability at the reattachment~\citep{navarro2005numerical,chuvakhov2017effect,roghelia2017experimental,re2017experimental}. In contrast, our analysis shows that the reattachment streaks in a cold-wall hypersonic compression ramp flow are caused by the baroclinic effects. These effects arise from the interactions of base flow density gradients in the thermal boundary layer with spanwise gradients of the incoming pressure perturbations and are a distinguishing feature of high-speed cold-wall compressible flows. We have also demonstrated that the spanwise scale selection can be attributed to the presence of flow perturbations in the separation bubble and in the reattaching shear layer. In particular, the weak amplification for small spanwise wavelengths results from the absence of perturbation dynamics within the bubble. In contrast, for perturbations with large spanwise length scales, amplification is weak because the perturbations in the separation bubble destructively interfere with the perturbations in the separated shear~layer.

The I/O approach provides a useful computational framework to quantify the spatial evolution of external perturbations in shock/boundary layer interactions. Improved understanding of amplification mechanisms can provide important physical insights about transition to turbulence. We expect that our work will motivate additional numerical and experimental studies that explore nonlinear aspects of transition in complex high speed flows and pave the way for the development of predictive transition models and effective flow control strategies.
 
\begin{figure}
\centering
\includegraphics[width=0.8\linewidth]{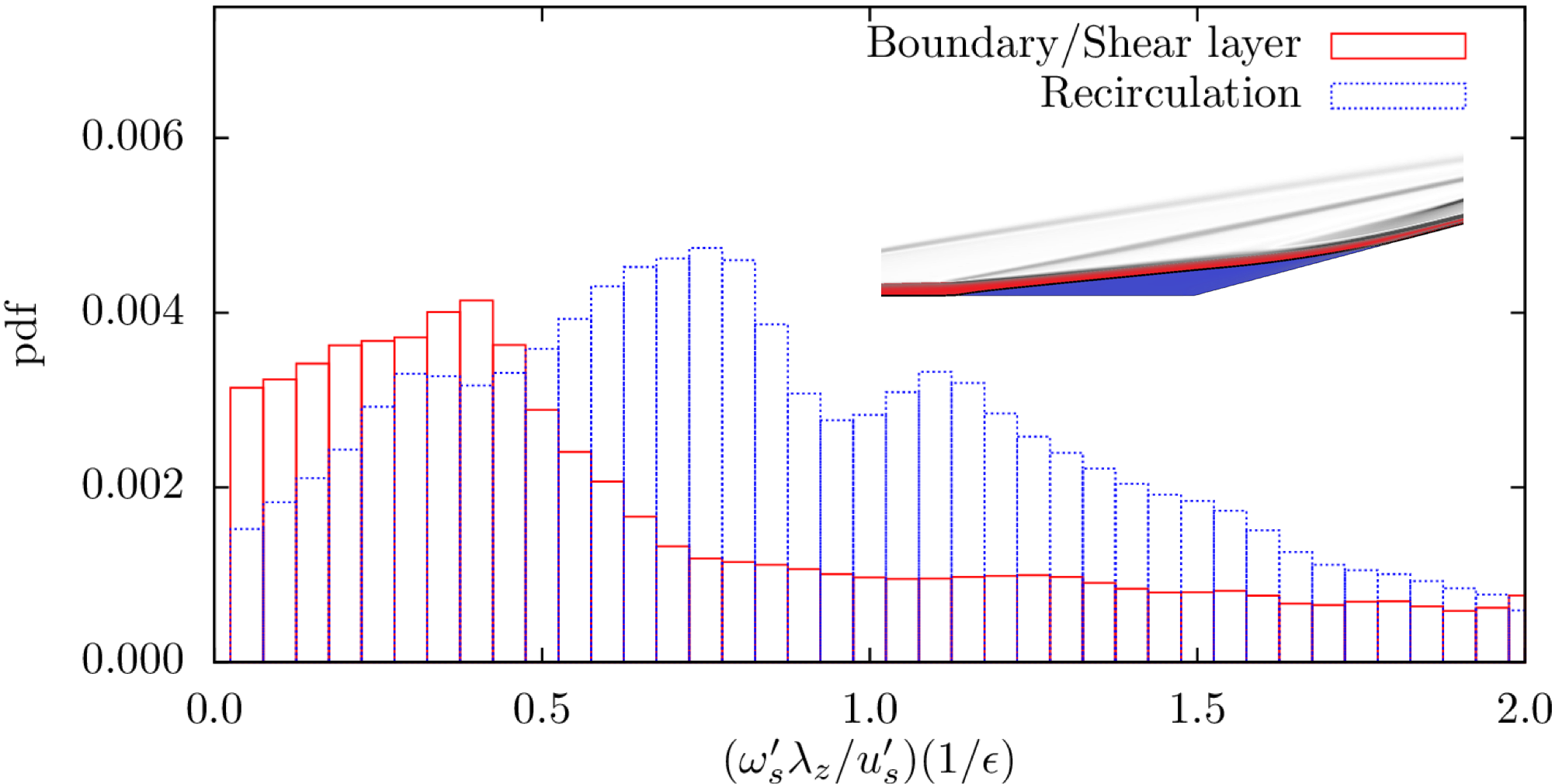}
\caption{Probability distribution function comparing relative size of streamwise vorticity $\omega'_{s}$ and velocity $u'_{s}$ perturbation components of the dominant output $\bphi_{1}$ for spanwise wavelength $\lambda_{z}=3$. The Inset shows regions corresponding to the recirculation bubble and the boundary/shear layer.}
\label{fig:pdf}
\end{figure}

\vspace*{-2ex}
\section*{Acknowledgements}

Financial support from the Air Force Office of Scientific Research (under award FA9550-18-1-0422) and the Office of Naval Research (under awards N00014-17-1-2496 and N00014-19-1-2037) is gratefully acknowledged. 
 
	%\newpage
	\vspace*{-2ex}
	\appendix
\section{Relative scaling of perturbation velocity components}
\label{appendixA}

Streak-like perturbations in attached boundary layers obey a scaling similar to the base flow quantities. Since the streamwise velocity perturbations are significantly larger than the wall-normal and spanwise perturbations, we have $u_{n,z}'/u_s' \approx \mathcal{O}(\epsilon)$ where $\epsilon = \bar{u}_{\eta}/\bar{u}_{\xi}$ is the small parameter and ($\eta,\xi$) denotes the wall aligned coordinate system. Here, we verify the validity of this relative scaling in the steady output mode associated with the compression ramp SBLI flow. 

Analysis is done for $Re_{L} = 3.7 \times 10^5$ and the dominant output at this flow condition corresponds to $\lambda_z = 3$. We study the relative scaling of the streamwise velocity $u'_s$ and vorticity $\omega'_s \DefinedAs \partial_n w' - \mri \beta u_n'$ perturbations associated with the dominant output by examining the quantity $\omega'_s \lambda_z \epsilon/u'_s$ at each spatial location in the flow-field. If a relative scaling described above holds, this quantity should be of $\mathcal{O}(1)$. Figure~\ref{fig:pdf} shows the probability distribution function (pdf) of this quantity, sampled inside and outside the recirculation region. In both regions, the quantity is indeed of $\mathcal{O}(1)$, confirming the scaling. We also observe that, when normalized with the streamwise component, the $\omega'_{s}$ is larger in the recirculation region than in the shear/boundary layer. 

\vspace*{-2ex}
\section{Spatial transport of streamwise vorticity}
\label{appendixB}
The steady inviscid streamwise vorticity equation in the ($s,n$) coordinates is given by
\begin{equation}
u_j \partial_j \omega_s = \omega_j \partial_j u_s - \omega_s \partial_j u_j + \dfrac{\partial_n \rho \partial_z p - \partial_z \rho \partial_n p}{\rho^{2}}  -  \dfrac{u_s \omega_n}{\mathcal{R}} \qquad j=s,n,z \label{oms_total}
\end{equation}
Linearization of Eq.~\eqref{oms_total} around base velocity ($\bar{u}_s$,0,0) and vorticity (0,0, $\bar{\omega}_z$) fields yields
\[
	\bar u_s \partial_s \omega'_s 
	\; = \; 
	(\bar \omega_z \partial_z u'_s  + \partial_s \bar u_s\,\omega'_s+  \partial_n \bar u_s\,\omega'_n) 
	\; - \; 
	\partial_s \bar u_s\,\omega'_s \; + \; 
	(\dfrac{\partial_n \bar  \rho}{\bar \rho^2} \, \partial_z p' - \dfrac{\partial_n \bar p}{\bar \rho^2} \, \partial_z \rho') 
	\; - \; \dfrac{\bar u_s}{\mathcal{R}}\,\omega'_n.   
\]
As shown in Appendix~\ref{appendixA}, for the dominant output we have $\partial_n \bar \rho \, \partial_{z} p' \gg  \partial_n \bar p \, \partial_z \rho'$ and $ \partial_{z} u'_{s} \gg \partial_s w' $. Since $\bar \omega_z  \DefinedAs - \partial_n \bar u_s - {\bar u_s}/{\mathcal{R}}$ and $\omega'_n  \DefinedAs - \partial_s w' + \partial_{z} u'_s $, for spanwise periodic perturbations with the spanwise wavenumber $\beta$ we obtain
\begin{align}
    \bar u_s \partial_s \omega'_s 
    \; \approx \; 
    -\partial_s w' \partial_n \bar u_s 
    \; - \; 
    2 \, \dfrac{\bar u_s}{\mathcal{R}} \, \mri \beta u'_s 
    \; + \; 
    \dfrac{\partial_n \bar \rho}{\bar \rho^2} \,  \mri \beta p'.
\end{align}

\vspace*{-3ex}
\section{Transport equation for Chu's compressible energy}
\label{appendixC}
Chu's compressible energy is determined by the quadratic form of the state $\B q$ of the linearized evolution model~\eqref{eq:SScomp},
\begin{align}
\Ece \;=\; \bphi^* {\B Q} \bphi \; = \; {\B q}^* {\B M} {\B q},
\end{align}
where the matrix $\B Q$ incorporates the quadrature weights as well as the diagonal transformation matrix that depends on the base flow quantities~\citep{hanifi1996transient}, and ${\B M} = {\B C}^* {\B Q} {\B C}$. By introducing a coordinate transformation 
	\[
	\tilde{\B q}
	\; = \;
	{ \B M }^{\frac{1}{2}} {\B q},
	\]
the state equation in~\eqref{eq:SScomp} takes the form,
	\begin{equation}
	\frac{\mrd}{\mrd t} \, \tilde{\mathbf{q}} 
	\; = \; 
	{ \B M }^{\frac{1}{2}} {\B A} { \B M }^{-\frac{1}{2}} \tilde{\B q} \; + \; { \B M }^{\frac{1}{2}} {\B B} \B d,
	\label{eq.qtilde}
	\end{equation}
and the square of the Euclidean norm of $\tilde{\mathbf{q}}$ gives Chu's compressible energy, $\Ece = \tilde{\mathbf{q}}^* \tilde{\mathbf{q}}$. Left multiplication of~\eqref{eq:chu_prod} with $\tilde{\mathbf{q}}^*$ yields Eq.~\eqref{eq:chu_prod}, where work done by external disturbances $\mathcal{F}$ is determined by the inner product of $\tilde{\mathbf{q}}$ with $\B M^{\frac{1}{2}}  \B B \B d$. Expressions for transport, source, production, and viscous terms in Eq.~\eqref{eq:chu_prod} are provided in~\citet[Eqs.\ ($16$) and ($17$)]{sidharth2018onset}.

	% \vspace*{-4ex}

\end{document}